\documentclass[runningheads,a4paper]{llncs}

\usepackage{amssymb}
\usepackage{graphicx}
\usepackage{amsfonts}
\usepackage{amsmath}
\usepackage[export]{adjustbox}
\usepackage[inline]{enumitem}
\usepackage{caption,subcaption}
\usepackage{url}
\urldef{\mailsa}\path|{jaakko.peltonen, ziyuan.lin}@aalto.fi|
\newcommand{\keywords}[1]{\par\addvspace\baselineskip
\noindent\keywordname\enspace\ignorespaces#1}

\begin{document}
\mainmatter 








\title{Peacock Bundles: Bundle Coloring for Graphs with Globality-Locality Trade-off}


\author{Jaakko Peltonen$^{1,2}$ \and Ziyuan Lin$^1$}
\authorrunning{Jaakko Peltonen and Ziyuan Lin}
\institute{$^1$Helsinki Institute for Information Technology HIIT, Department of Computer Science, Aalto University, Finland\\
$^2$School of Information Sciences, University of Tampere, Finland\\
\mailsa}

\toctitle{Peacock Bundles: Bundle Coloring for Graphs with Globality-Locality Trade-off}
\titlerunning{Peacock Bundles: Bundle Coloring for Graphs with Globality-Locality Trade-off}
\tocauthor{Jaakko Peltonen and Ziyuan Lin}
\maketitle



\begin{abstract}
Bundling of graph edges (node-to-node connections) is a common technique to enhance 
visibility of overall trends in the edge structure of a large graph layout, and a large variety 
of bundling algorithms have been proposed. However, with strong bundling, it becomes hard to 
identify origins and destinations of individual edges. 
We propose a solution: we optimize edge coloring to differentiate bundled edges. 
We quantify strength of bundling in a flexible pairwise fashion between edges, 
and among bundled edges, we quantify how dissimilar their colors should be
by dissimilarity of their origins and destinations. 
We solve the resulting nonlinear optimization, which is also interpretable as a novel
dimensionality reduction task.
In large graphs the necessary compromise is whether to differentiate colors sharply 
between locally occurring strongly bundled edges (``local bundles''), or also between 
the weakly bundled edges occurring globally over the graph (``global bundles'');
we allow a user-set global-local tradeoff.
We call the technique ``peacock bundles''. 
Experiments show the coloring clearly enhances comprehensibility of graph layouts with edge bundling.
\end{abstract} 












\section{Introduction}

Graphs are a prominent type of data in visual analytics.  Prominent
graph types include for instance hyperlinks of webpages, social
networks, citation networks between publications, interaction networks
between genes, variable dependency networks of probabilistic graphical
models, message citations and replies in discussion forums, traces of
eye fixations, and many others.
2D or 3D visualization of graphs is a common need in data
analysis systems. 
%
If node coordinates are not available from the data,
several node layout methods have been developed, from constrained
layouts such as circular layouts ordered by node degree to
unconstrained layouts optimized by various criteria; the latter
methods can be based on the node and edge set (node adjacency matrix)
alone, or can make use of multivariate node and edge features, 
typically aiming to reduce edge crossings and place nodes close-by if
they are similar by some criterion.

In layouts with numerous edges it may be hard to see trends in
node-to-node connections. Edge bundling 
draws multiple edges as curves that are close-by and
parallel for at least part of their length. Bundling simplifies the
appearance of the graph, and bundles 
also
summarize
connection trends between areas of the layout.
However,
when edges are drawn close
together, the ability to visually follow edges and
discover 
their start and end points is lost. 
Interactive systems 
\cite{grossman2005bubble}
can allow inspection of edges,
but inspecting numerous edges is laborious.

\begin{figure}[!t]
\centering
\includegraphics[width=0.25\textwidth]{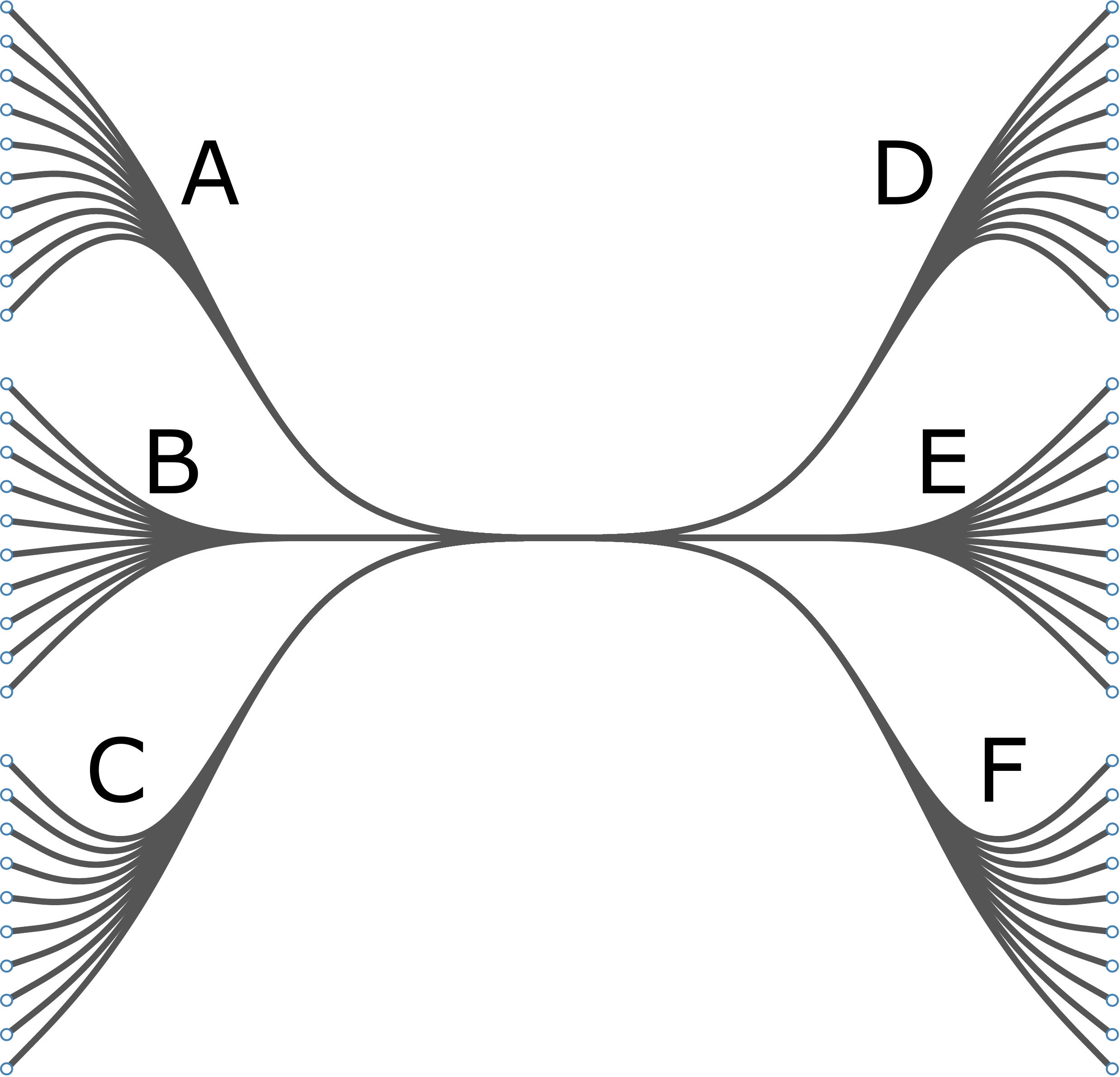}
\includegraphics[width=0.25\textwidth]{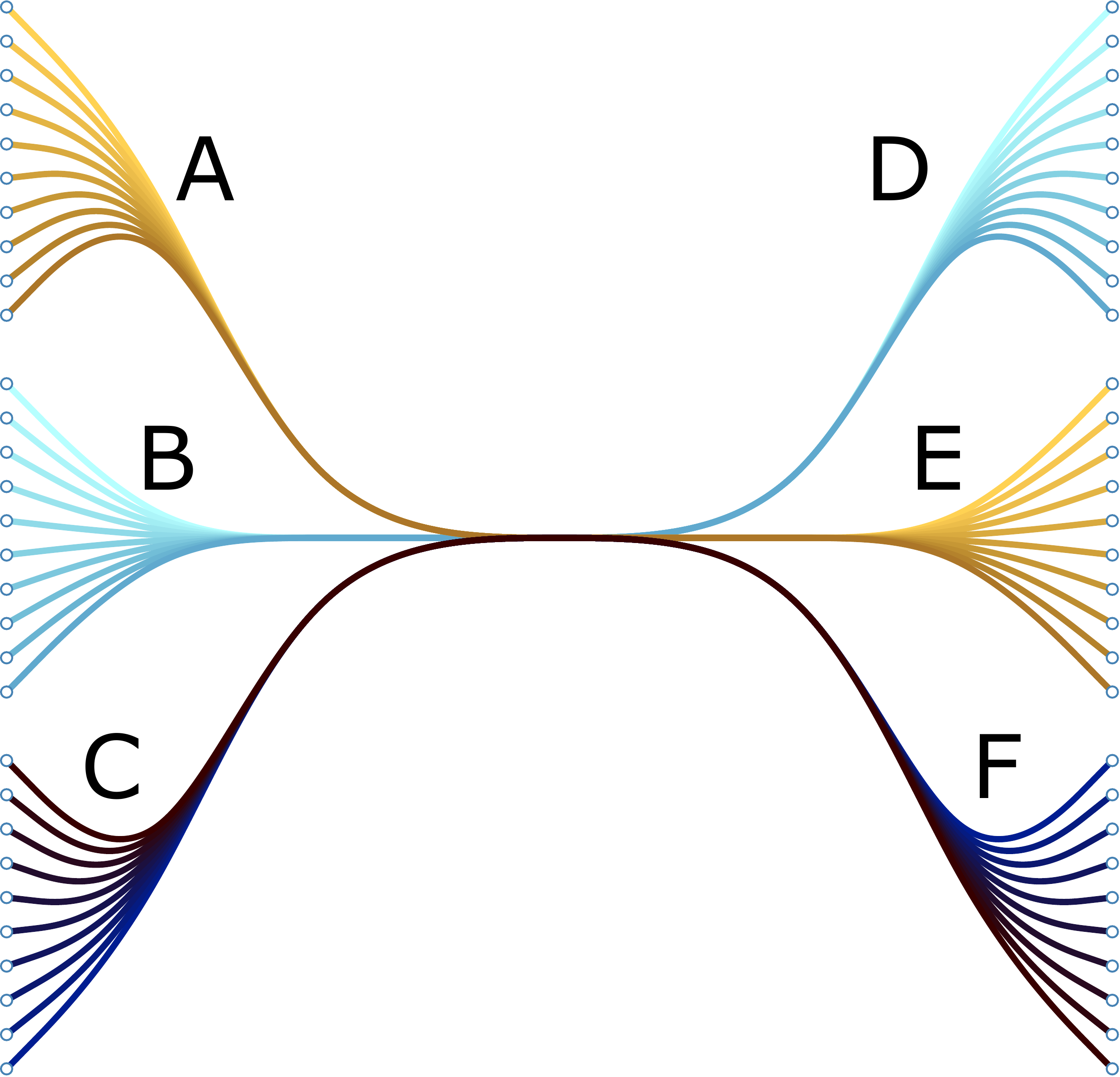}
\includegraphics[width=0.45\textwidth]{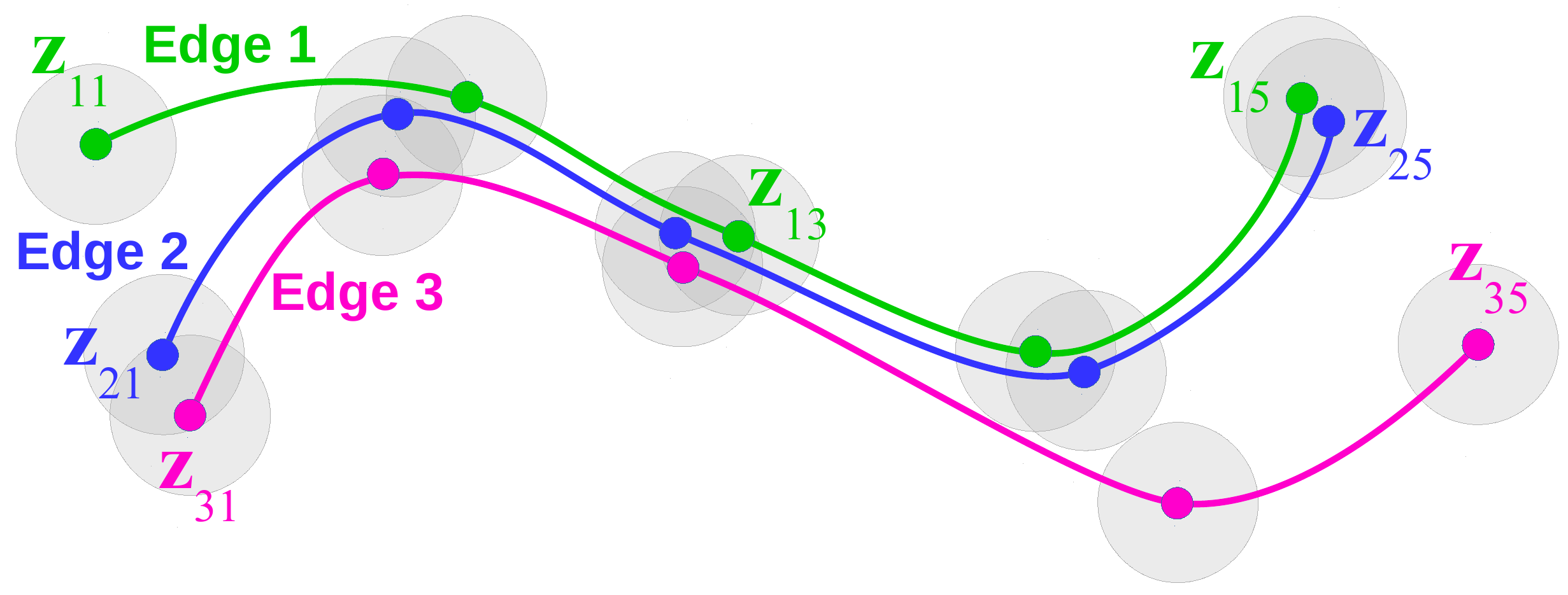}

\caption{
Illustration of peacock bundle coloring.
\textbf{Left:} A graph with node groups A-F, drawn with hierarchical edge bundling. 
With plain gray coloring finding the connecting vertex pairs is not possible.
\textbf{Middle:} Peacock bundle coloring 
reveals
that nodes in group A connect to nodes in group E in order, and similarly B to D in order, 
and C to F in reverse order. The connections are easily seen from 
the optimized coloring produced by our Peacock Bundles method: bundled
edges traveling from and to close-by nodes get close-by colors. 
\textbf{Right:} Pairwise bundling detection as described in Section \ref{sec:bundling_detection},
for three edges, control points $\mathbf{z}_{ij}$ shown as 
circles, distance threshold $T$ as the radius of light gray circles (small threshold used for 
illustration). Control points $\mathbf{z}_{12},\ldots,\mathbf{z}_{15}$ of edge 1 are near control 
points of edge 2, but only $\mathbf{z}_{13}$ is near a control point of edge 3. If, e.g., 
$K_{ij}=2$ nearby control points are required between edges, edge 1 is considered bundled with 
edge 2 but not edge 3; edge 2 is considered bundled with edges 1 and 3, and edge 3 with edge 2 
but not edge 1. Since edges 1 and 3 are not considered bundled they could be assigned a similar 
color.
}
\label{fig:bundling_detection}
\end{figure}

Comprehensibility of edges can be enhanced by distinguishing them
by visual properties,
such as line style, line width,
markers along the curve, or color. 
Following an edge by
its color can allow an analyst to see where each edge goes, but poorly
assigned colors can make this task hard to do at a glance.  \textbf{We
present a machine learning method that optimizes edge
colors in graphs with edge bundling, to keep bundled edges maximally
distinguishable.}  We focus on edge color as it has several
degrees of freedom suitable for optimization (up to
three continuous-valued color channels if using RGB color space), 
but 
our method
is easily
applicable to 
other continuous-valued edge properties.
We call our solution \textbf{peacock bundles} as it is inspired by
the plumage of a peacock; our method results in a
fan of colors, reminiscent of a peacock tail, at
fan-in locations of edges arriving into a bundle and  
fan-out locations of edges departing from a bundle.
Figure \ref{fig:bundling_detection} (middle) illustrates the concept
and how it can help follow edges. 
We next review related works and then present the method and experiments.

\section{Background: node layout, edge bundling, and coloring}

Node layouts of graphs have been optimized by many approaches, see
\cite{gibson13} for a survey. Our approach is not specific to
any node layout approach and can be run for any resulting layout.
Several methods have been proposed for edge bundling 
\cite{cui08,holten06,dwyer06,hurter12a,ersoy11,pupyrev10,luo12,gansner11}.
For example, Cui et al. \cite{cui08} generate a mesh covering the graph on the display 
based on node positions and edge distribution. 
The mesh helps cluster edges spatially; edges within a cluster are bundled.
Hierarchical Edge Bundling \cite{holten06} 
embeds a tree representation for 
data 
with hierarchy onto the 2D display. 
Tree nodes are used as spline control points for edges; 
bundles come from reusing control points.
See Zhou \cite{zhou13} for a recent review and taxonomy.  

Unlike node layout and edge bundling, relatively little attention has been
paid to practical edge coloring; while graph theory papers
exist about the ``edge coloring problem'' of 
setting
distinct colors to adjacent edges with a minimum number of
colors, that combinatorial problem does not reflect 
real-life
graph visual analytics where a continuous edge color space exists
and the task is 
to 
set
colors to be
informative about graph properties.
Simple coloring approaches exist. A naive coloring 
sets a random color to each edge: such coloring 
is unrelated
to spatial positions of nodes and edges and 
is chaotic, making
it hard to 
grasp an overview of edge
origins and destinations 
at a glance. Edge colors are
sometimes reserved to show discrete or multivariate annotations 
such as edge strengths; such coloring relies on external data
and may not help gain an overview of the graph layout itself. A
simple layout-driven solution is to color each edge
by
onscreen position of the start or end node. If edges
have been clustered by some method, one often sets the same color
to the whole cluster \cite{ersoy11,telea2010image}; this simplifies coloring, but
prevents telling apart origins and destinations of
individual edges. 

Hu and Shi \cite{hu2014coloring} create edge colorings
with a maximal distinguishability motivation related to ours,  
but
their method does not consider actual edge bundling
and operates on the original graph; we 
operate on 
bundled graphs and quantify edge bundling.
Also,
instead of 
only
using
binary detection of 
bundled edge pairs (a hard criterion whether two edges are bundled) 
we differentiate all edges, emphasizing
each pair by a weight that is high between strongly bundled local edges
and smaller between others, with a user-set global-local tradeoff. 
Lastly, their method tries to set maximally distinct colors between all bundled
edges, needing 
harder compromises for larger bundles: 
we quantify which bundled edges need the most distinct colors by comparing 
their origin and destination coordinates in the layout, and thus can devote 
color resources efficiently even in large graphs.


Our algorithm 
is related
to nonlinear dimensionality reduction.
Although dimensionality reduction has been used in colorization 
for other domains \cite{daniels2010interactive,casaca2012colorization},
to our knowledge ours is the first method to optimize 
local graph coloring with edge bundling as a 
dimensionality reduction task.

\section{The method: Peacock bundles}

Bundle coloring has several challenges. \textbf{1.}
Efficient coloring should depend
not only on high-dimensional graph properties on the low-dimensional 
graph layout: if two edges are spatially distinct they do not need 
different colors.
\textbf{2.} \emph{Bundles are typically
not clearly defined}: the curve corresponding to an individual edge may
become \emph{locally} bundled with several other edges at different places
along the curve between its start and end node, and edges cannot be
cleanly separated into groups that would correspond to some \emph{globally}
nonoverlapping bundles. Solutions requiring nonoverlapping bundles would be suboptimal: 
they would either not be applicable to real-life edge-bundled graphs or would need to 
artificially approximate the bundle structure of such graphs as nonoverlapping subsets.
\textbf{3.} The solution should scale up to large graphs with large bundles.
In large bundles it is typically not feasible to assign strongly distinct colors
between all edges; it is then crucial to quantify how to make the compromise, that is,
which edges should have the most distinct colors within the bundle.

Our coloring solution neatly solves these
challenges, by posing the coloring as an optimization task defined
based on local bundling between \emph{each individual pair of edges}. 
Our solution is applicable to all graphs and takes into account the full 
bundle structure in a graph layout without approximations.
For any two edges it is easy to define whether their curves
are bundled (close enough and parallel) for some part of
their length, without requiring a notion of a globally defined bundle;
we optimize the coloring to \emph{tell each edge apart from the ones it
has been bundled with}.
Such optimization makes maximally efficient use
of the colors: two edges need distinct colors only if they are bundled
together, whereas two edges that are not bundled can share the same
color or very similar colors. Moreover, even between two bundled edges,
how distinct their colors need to be can be quantified in a natural way
based on the node layout: the more dissimilar their origins and destinations 
are,
the more dissimilar their colors should be.
Differentiating origins and destinations helps analysts assuming the node layout is meaningful.  
%
Computation of peacock bundles requires two steps:
\begin{enumerate}
\item \emph{Detection of which pairs of edges are bundled together} at
  some location along their curve. We solve this by a 
  well-defined 
  closeness threshold of consecutive curve segments. 
  An edge may participate in multiple bundles along
  its curve.
\item \emph{Definition of the color optimization task}. We formalize
  the color assignment task as a dimensionality reduction task from
  two input matrices, a pairwise edge-to-edge bundling matrix and a
  dissimilarity matrix that quantifies how dissimilar colors of bundled edges should be,
  to a continuous-valued low-dimensional
  colorspace, which can be one-dimensional (1D) to achieve a
  color gradient, or 2D or 3D for greater variety. 
  (Properties like width or continuous line-style attributes could be included  
  in a higher than 3D output space;
  here we use color only.)
  We define the color assignment as an
  edge dissimilarity preservation task: colors are optimized
  to preserve spatial dissimilarities of start and end nodes among
  each pair of bundled edges, 
whereas 
no constraint is placed between
  colors of non-bundled edge pairs.
\end{enumerate}

Peacock bundle coloring can be integrated into 
edge bundling algorithms, but can be also run as standalone
postprocessing for graphs with edge bundling,
regardless of which algorithms yielded the node layout and
edge curves. Peacock bundles
optimize colors taking both the graph and its visualization
(node and edge layout) into account:
color separation needs to be emphasized only for edges that appear
spatially bundled.
We demonstrate the result on several graphs with different node
layouts and a popular edge bundling technique.

\subsection{Detection of bundled pairs of edges}
\label{sec:bundling_detection}

Let the graph contain $M$ edges $i=1,\ldots,M$, each represented by a
curve. If the curves are spline curves, let each curve be generated
by $C_i$ control points; if the curves are piecewise linear, let
each curve be divided into $C_i$ segments represented e.g. by the
midpoint of a segment.  
For brevity we use the terminology of control
points in the following, but the algorithm can be used just as well
for other definitions of a curve, such as midpoints of piecewise linear
curves
or equidistributed points on the curves if getting such locations is convenient.

Let $B_{ij}$ be a variable in $[0,1]$ denoting whether edge $i$ is bundled
together with edge $j$.  
If the edge bundling has been created by an algorithm that explicitly
defines bundle memberships for edges, $B_{ij}$ can simply be set to 1 
for edges assigned to the same bundle and zero otherwise. However,
for several situations this is insufficient: i) sometimes the bundling
algorithm is not available or the bundling has e.g. been created interactively;
ii) some bundling algorithms only e.g. attract edge segments and do not define
which edges are bundled; iii) an edge may be close to several different other
edges, so that no single bundle membership is sufficient to describe its
relationship to other edges. For these reasons we provide a way to
define pairwise edge bundling variables $B_{ij}$ that does not require
availability of any previous bundling algorithm.

We set $B_{ij}=1$ if at least
$K_{ij}$ consecutive control points of edge $i$ are each close enough
to one or more control points of edge $j$.  Intuitively, if several
consecutive control points of edge $i$ are close to edge $j$, the
edges travel close and parallel (as a bundle) at least between those
control points. 
Since our choice of control points does not allow the curves to change drastically between two consecutive control points,
the defined $B_{ij}$ is stable when the control point densities between curve $i$ and curve $j$ do not differ too much.
In practice, 
we set $K_{ij}$ to an integer at least 1, 
separately for each pair of edges,
as a fraction of the number of available control points
as detailed later in this section.

Formally, for edge $i$ denote the on-screen coordinates of the $C_i$
control points by $\mathbf{z}_{i1},\ldots,\mathbf{z}_{iC_i}$, and
similarly for edge $j$. Let $d(\cdot,\cdot)$ denote the Euclidean
distance between two control points, and let $T$ be a distance
threshold. Then
\begin{equation}
B_{ij}=\max_{r_0=1,\ldots,C_i-K_{ij}+1} \prod_{r=r_0}^{r_0+K_{ij}-1} 1(\min_{s=1,\ldots,C_j} d(\mathbf{z}_{ir},\mathbf{z}_{js})\le T)
\label{eq:thrsh-cp-int}
\end{equation}
where $r=r_0,\ldots,r_0+K_{ij}-1$ are indices of 
consecutive control points in edge $i$. The term $1(\cdot)$ 
is 1 if the statement inside is true and zero otherwise: that is, the term is 1 if the
$r$th control point of edge $i$ is close to edge $j$ (to some control point $s$ of edge $j$). 
The whole product term is 1 if the $K_{ij}$ consecutive control points of $i$ from $r_0$ onwards are all close to edge $j$. Finally, the whole term $B_{ij}$ is 1 if edge $i$ has
$K_{ij}$ consecutive points (from any $r_0$ onwards) that are all close to edge $j$.
Figure \ref{fig:bundling_detection} (right) illustrates the pairwise
bundling detection.




The distance threshold $T$ should be set to a value
below which line segments appear very similar; a rule of thumb
is to set $T$ to a fraction of the total diameter (or larger 
dimension) of the screen area of the graph. Similarly, 
a convenient way to set the required number of close-by control points 
$K_{ij}$ is to set it to a fraction of the 
maximum number of control points in the two edges, requiring at 
least 1 control point, 
so that for each pair of edges $i$ and $j$ we set 
$K_{ij}=\max(1,\lfloor\max(C_i,C_j)K_{min}\rfloor)$ where $K_{min}\in (0,1]$ is
the desired fraction.

Detected pairwise bundles match ground truth 
in all simple examples we tried (e.g. Fig.~\ref{fig:bundling_detection} left); 
in experiments of Section \ref{sec:experiments} where no ground truth is 
available the bundling is visually good; edges bundled with 
any edge of interest can be interactively checked at
\url{http://ziyuang.github.io/peacock-examples/}.




\subsection{Optimization of edge colors by dimensionality reduction}

Our coloring 
is based on dimensionality reduction of bundled
edges from an original dissimilarity (distance) matrix to a color
space; we thus need to define how dissimilar two bundled edges
are. We aim to help analysts differentiate where in the graph
layout each edge goes; we thus use the node locations of edges to
define the similarity.
Denote the two on-screen node layout coordinates of edge $i$ by 
$\mathbf{v}^1_i$ and $\mathbf{v}^2_i$. We first define 
\begin{equation}
d^{\textrm{original}}_{ij} = \min(\|\mathbf{v}^1_i-\mathbf{v}^1_j\|+\|\mathbf{v}^2_i-\mathbf{v}^2_j\|,\; \|\mathbf{v}^1_i-\mathbf{v}^2_j\|+\|\mathbf{v}^2_i-\mathbf{v}^1_j\|) \;.
\label{eq:dist-orig}
\end{equation}
Denote the set of $p$ features for edge $i$ as a vector
$\mathbf{x}_i=[x_{i1},\ldots,x_{ip}]$, and denote the low-dimensional
output features for edge $i$ as a vector
$\mathbf{y}_i=[y_{i1},\ldots,y_{iq}]$ where $q\in\{1,2,3\}$ is the output
dimensionality.
We define the dimensionality reduction task as minimizing the difference between
the endpoint dissimilarity of bundled edges and dissimilarity of their optimized
colors. 
This yields the cost function
\begin{equation}
\min_{\{\mathbf{y}_1,\ldots,\mathbf{y}_M\}} \sum_i \sum_j B_{ij}(d^{original}_{ij}-d^{out}(\mathbf{y}_i,\mathbf{y}_j))^2
\label{eq:optimization_cost}
\end{equation}
where $d^{out}(\mathbf{y}_i,\mathbf{y}_j)$ is
the Euclidean distance between the output features. The terms $B_{ij}$ are large
for only those pairs of edges that are bundled, thus minimizing the cost
assigns colors to preserve dissimilarity within bundled edges, but allows
freedom of color assignment between non-bundled edges.
The cost encapsulates that 
greater difference of edge destinations should 
yield greater color difference, and that color differentiation is most needed
for strongly bundled edges.
While alternative formulations are possible, (\ref{eq:optimization_cost}) 
is simple and works well.

\textbf{From local to global color differentiation.}
The weights $B_{ij}$ detect edges according to thresholds $T$ and $K_{ij}$.
Some edge pairs that fail the detection might still 
visually appear nearly bundled: instead of differentiating only 
within detected bundles, it is meaningful 
to 
differentiate
other edges too.
The simplest way is to encode a \emph{tradeoff} between local 
(within-bundle) and global differentiation
in the $B_{ij}$: we set $B_{ij}=1$ if edges $i$ and $j$ are bundled,
otherwise $B_{ij}=\epsilon$ where $\epsilon\in[0,1]$ is a user-set parameter
for the preferred global-local tradeoff.
When $0<\epsilon<1$, the cost emphasizes achieving desired color
differences between bundled edges (where $B_{ij}=1$) according to their
dissimilarity of origins and destinations, but also aims to achieve color differences between
other edges ($B_{ij}=\epsilon$) according to the same dissimilarity.
As the optimization is based on desired dissimilarities between edges, 
it intelligently optimizes colors even when all edge pairs can have nonzero weight:
$\epsilon=0$ means a pure local coloring where only bundled edge pairs matter, 
and $\epsilon=1$ means a pure global coloring that aims to show 
dissimilarity of origin and destination for all edges regardless of bundling.
In our tests coloring changes gradually with respect to $\epsilon$.
In experiments, when emphasizing local color differences, 
we set $\epsilon=0.001$ which achieved local differentiation
and formed color gradients for bundles in most cases.

A way to set a more nuanced tradeoff is to 
run edge detection with multiple settings and set weaker $B_{ij}$ for 
edges detected with weaker thresholds; in practice the above 
simple tradeoff already worked well. 


\textbf{Relationship to nonlinear multidimensional scaling.}
Interestingly, minimizing (\ref{eq:optimization_cost}) 
can be seen as a specialized weighted form of
\emph{nonlinear multidimensional scaling}, with several differences: unlike traditional 
multidimensional scaling we treat edges (not data items or nodes) as input items
whose dissimilarities are preserved; our output is not a spatial layout 
but a color scheme; and most importantly, the cost function does not aim to 
preserve all ``distances'' but weights each pairwise distance according
to how strongly that pair of edges is bundled.
The theoretical connection lets us make use of  
optimization approaches previously developed for multidimensional scaling, here 
we choose to use the popular stress majorization algorithm (SMACOF) \cite{borg05} to 
minimize the cost function.



\textbf{Color range normalization.}
After optimization, output features $\mathbf{y}_i$ of each edge
must be normalized to the range of the color channels (or positions along a color gradient).
Simple ideas like applying an affine transform to the output matrix $Y=(\mathbf{y}_1,\ldots,\mathbf{y}_M)$ 
would give different amounts of color space to different bundles, thus
colors within bundles would not be 
well
differentiated.
We propose a normalization to maximally distinguish edges within each bundle.
Let $Col$ denote the color matrix to be obtained from normalization.
For each $\mathbf{y}_i$, let $\{\mathbf{y}_{i_l}\}_{l=1}^{M_i}$ be the set of output features 
where each edge $l$ is bundled with edge $i$.
We assemble $\mathbf{y}_i$ and $\{\mathbf{y}_{i_l}\}_{l=1}^{M_i}$ into a matrix 
$Y^i=(\mathbf{y}_i,\mathbf{y}_{i_1},\ldots,\mathbf{y}_{i_{M_i}})$, 
affinely transform $Y^i$ to $\tilde{Y}^i=(\tilde{\mathbf{y}}_i,\tilde{\mathbf{y}}_{i_1},\ldots,\tilde{\mathbf{y}}_{i_{M_i}})$ so that each entry in $\tilde{Y}^i$
is within the allowed range (say, $[0,1]$), then set $Col_i$, the $i$-th column of $Col$
(color vector for edge $i$) as $\tilde{\mathbf{y}}_i$.
This normalization expands the color range within bundles.

\textbf{Where to show colors.}
The optimized colors can be shown along the whole edge, or
at ``fan-in'' segments where the edge enters a bundle and
``fan-out'' segments where it departs a bundle. Edge
$i$ is bundled with $j$ if several consecutive curve segments of $i$
are close to $j$; the last segment before the close-by ones is the
fan-in segment; the first segment after the close-by ones is the
fan-out segment. In experiments we show color along the
whole edge for simplicity.
Note that, as with any edge coloring, colors of close-by 
edges may perceptually blend, but our optimized colors 
then remain visible at fan-in and fan-out locations.


\begin{figure*}[!t]
\centering
\includegraphics[width=0.4\textwidth]{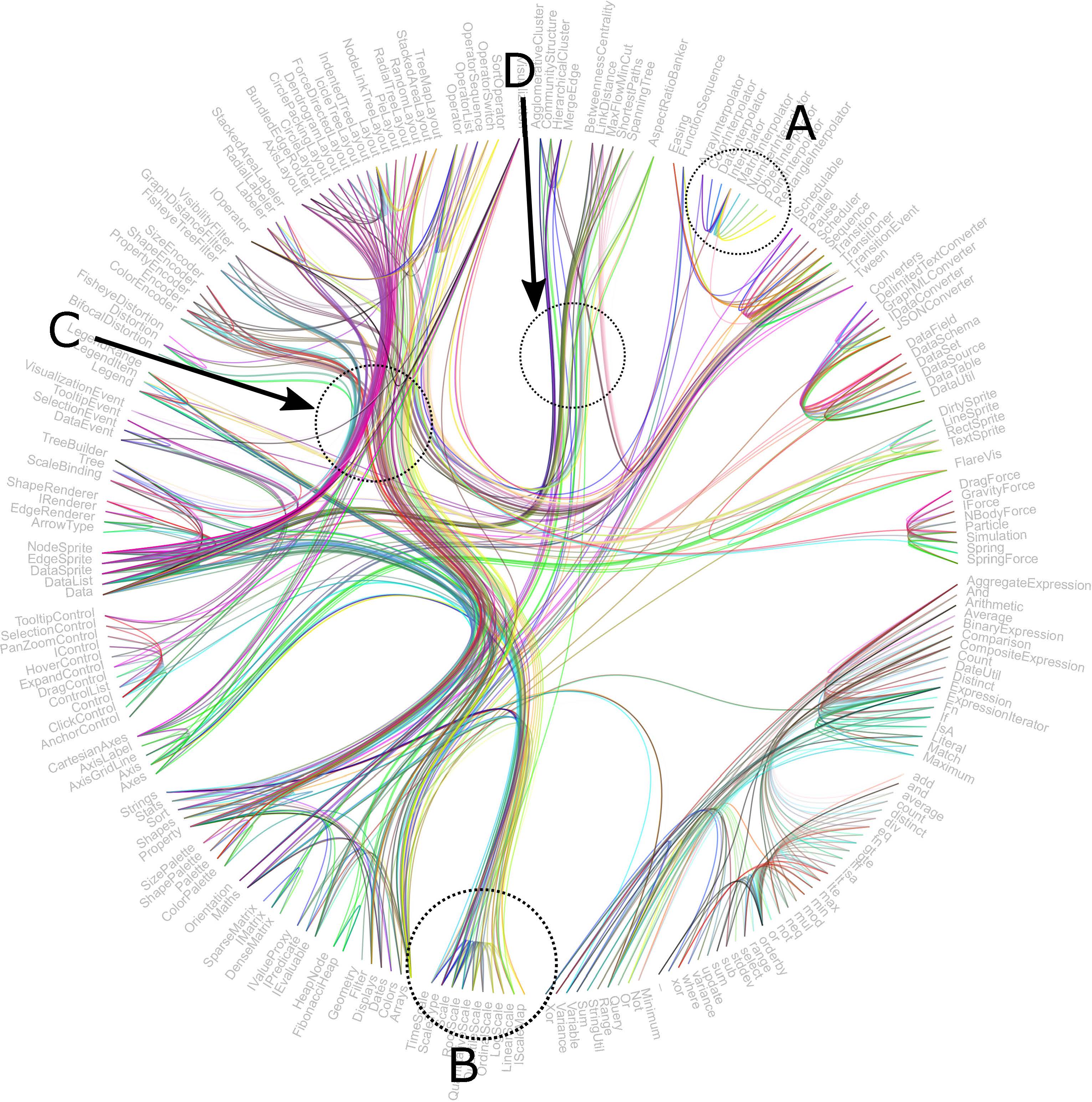}
\makebox[0.4\textwidth][c]{\includegraphics[width=0.35\textwidth]{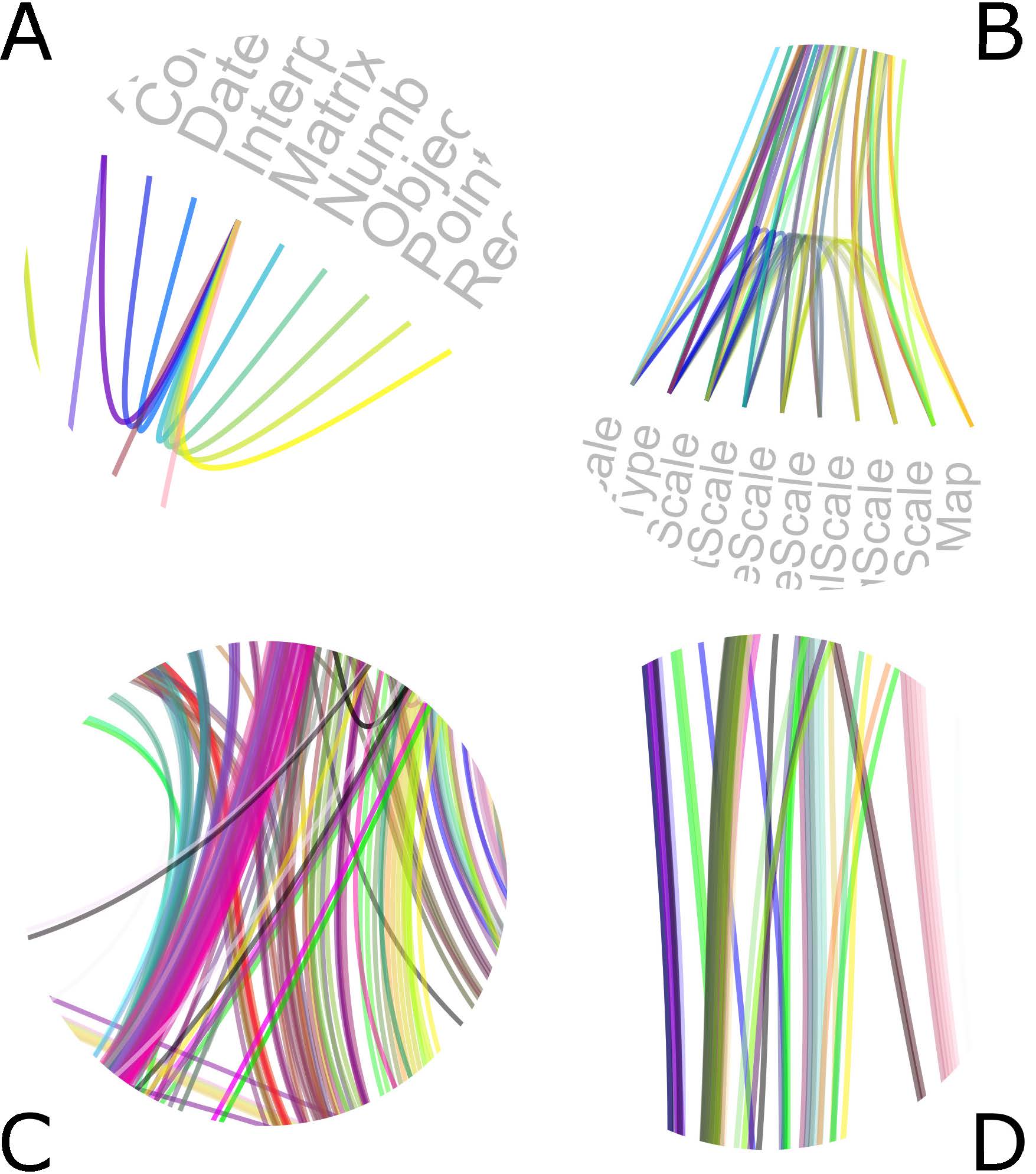}}
\includegraphics[width=0.4\textwidth]{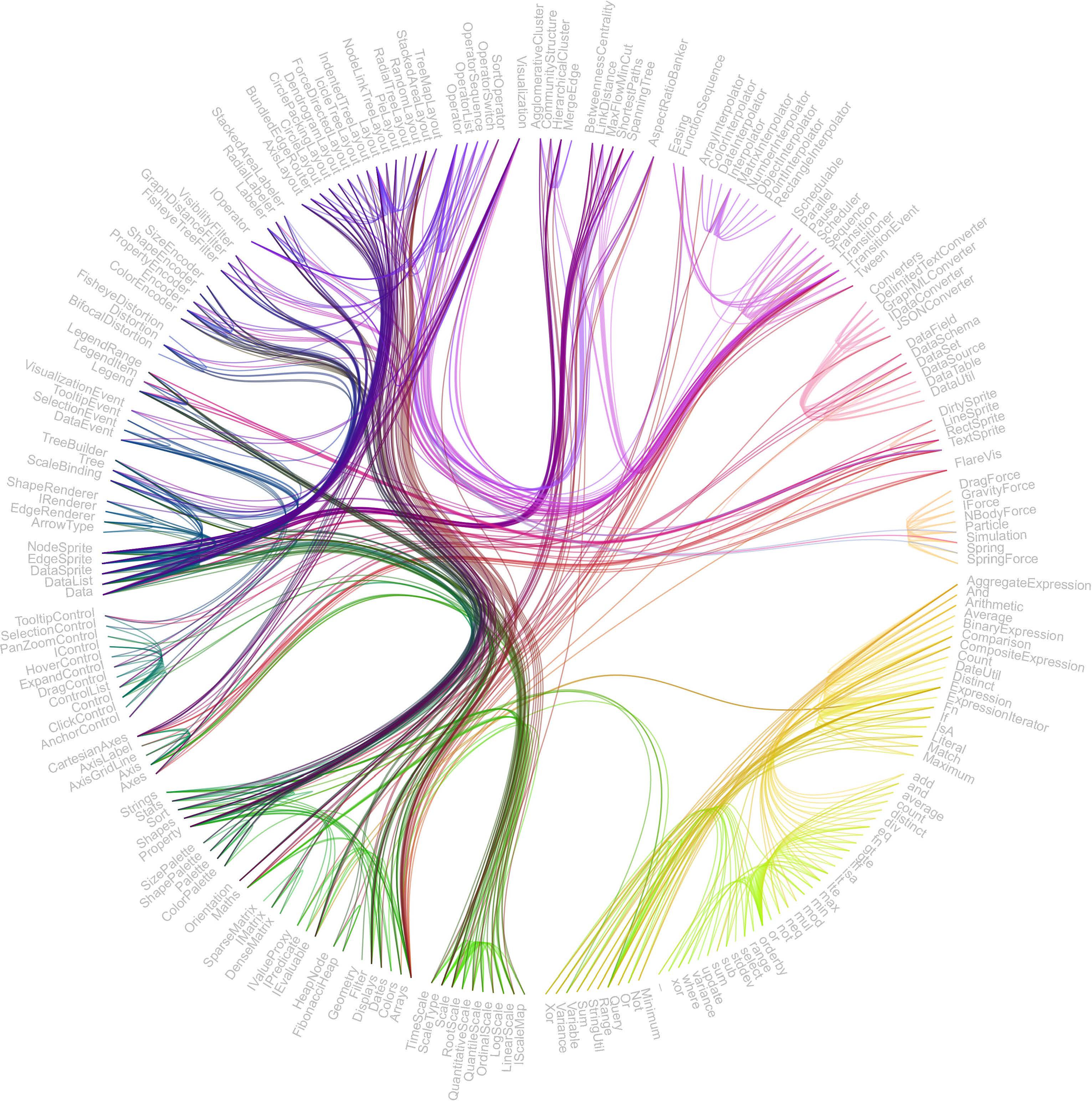}
\includegraphics[width=0.4\textwidth]{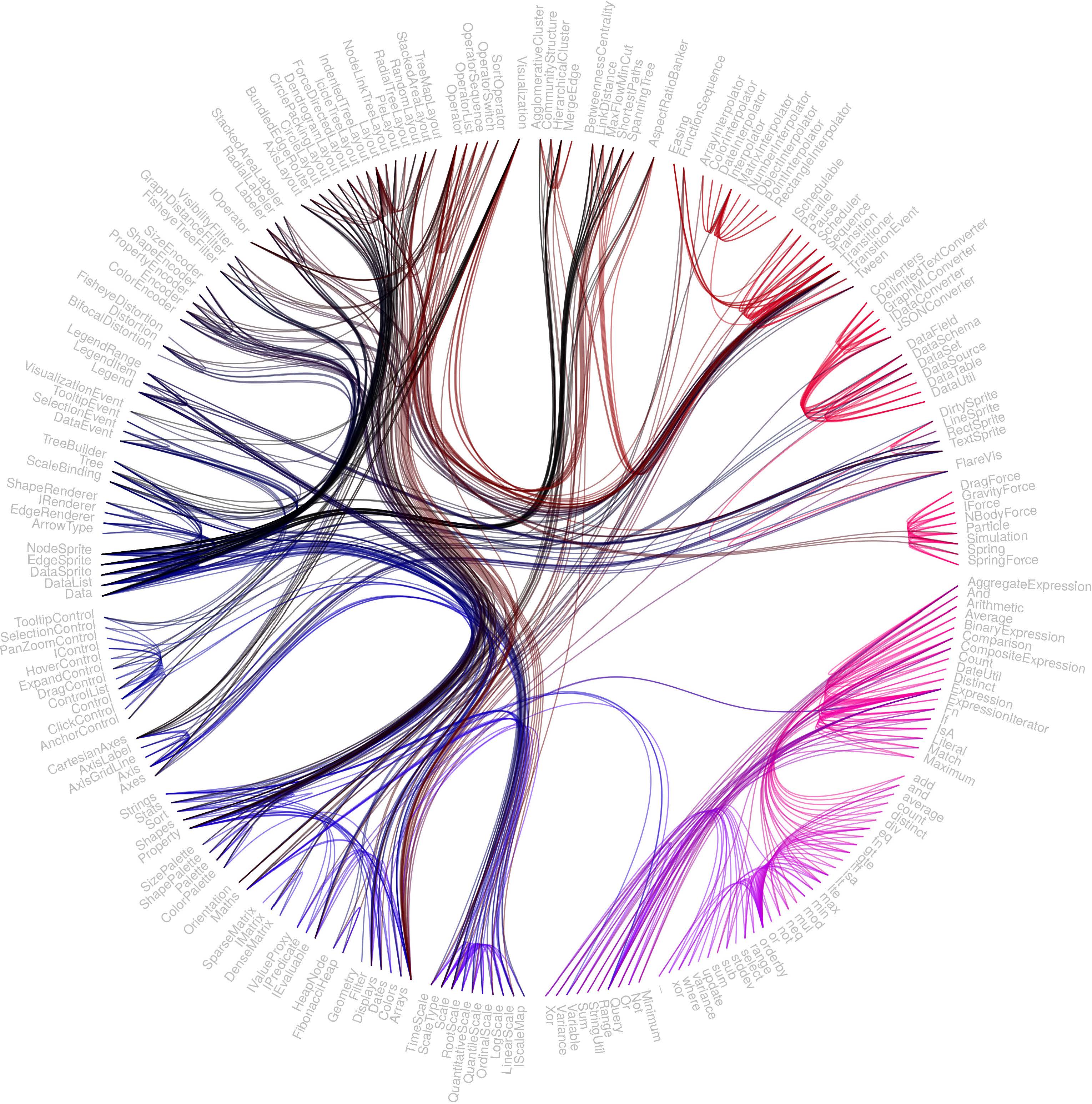}
\caption{Colorings for the graph ``radial''. \textbf{Top left}: the coloring from Peacock with $\epsilon=0.001$. 
\textbf{Top right}: zoomed-in versions of the parts within dashed-line circles 
in the top-left figure as examples of local coloring. In the four zoomed-in parts, 
colors show a linear gradient and vary in yellow-red-blue, thus
1) local colors are differentiated, and 
2) they span roughly the same full color range. 
The local colors also help follow edges at bottom right of the graph, where colors are homogeneous in the baseline coloring.
\textbf{Bottom left}: coloring from Peacock, $\epsilon=1$. 
The bundles are colored into 3 parts: the blue-ish upper half, green-ish lower-left part, 
and yellow-ish lower-right part. There are also red bundles joining the blue and green parts, differentiating itself from other bundles.
\textbf{Bottom right}: the baseline coloring. The bundles are colored into the red-ish upper half
and blue-ish lower half. Compared with the coloring with $\epsilon=1$ from Peacock, the bundle from left to right, and the bundle at the top-left corner are less distinguishable.
}
\label{fig:circular}
\end{figure*}

\section{Experiments}
\label{sec:experiments}

\begin{figure}[!t]
\centering
\begin{subfigure}[t]{\textwidth}
\includegraphics[width=0.49\textwidth]{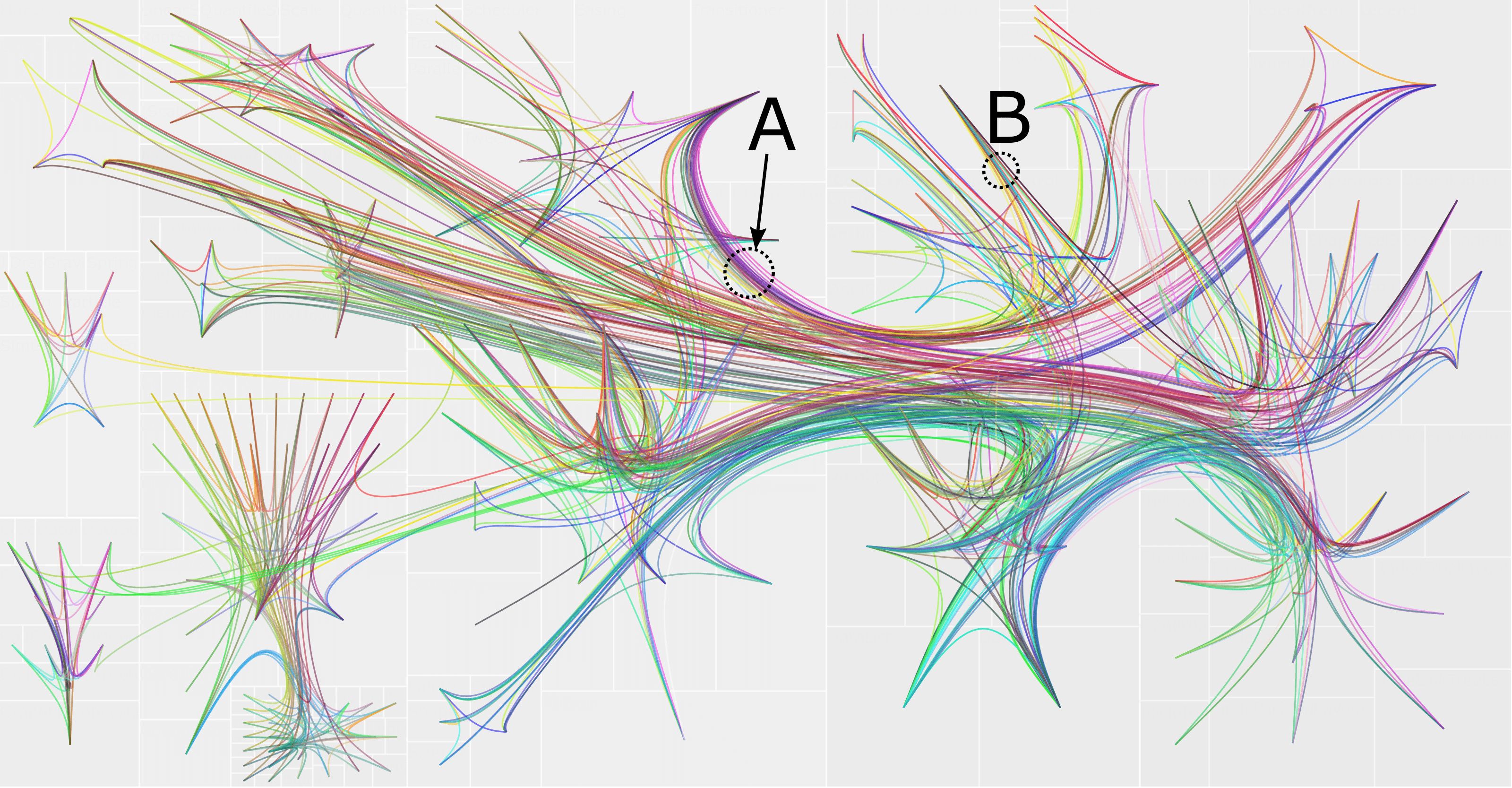}
\includegraphics[width=0.49\textwidth]{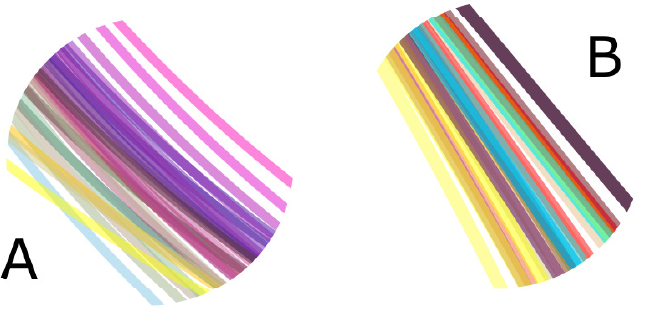}
\includegraphics[width=0.49\textwidth]{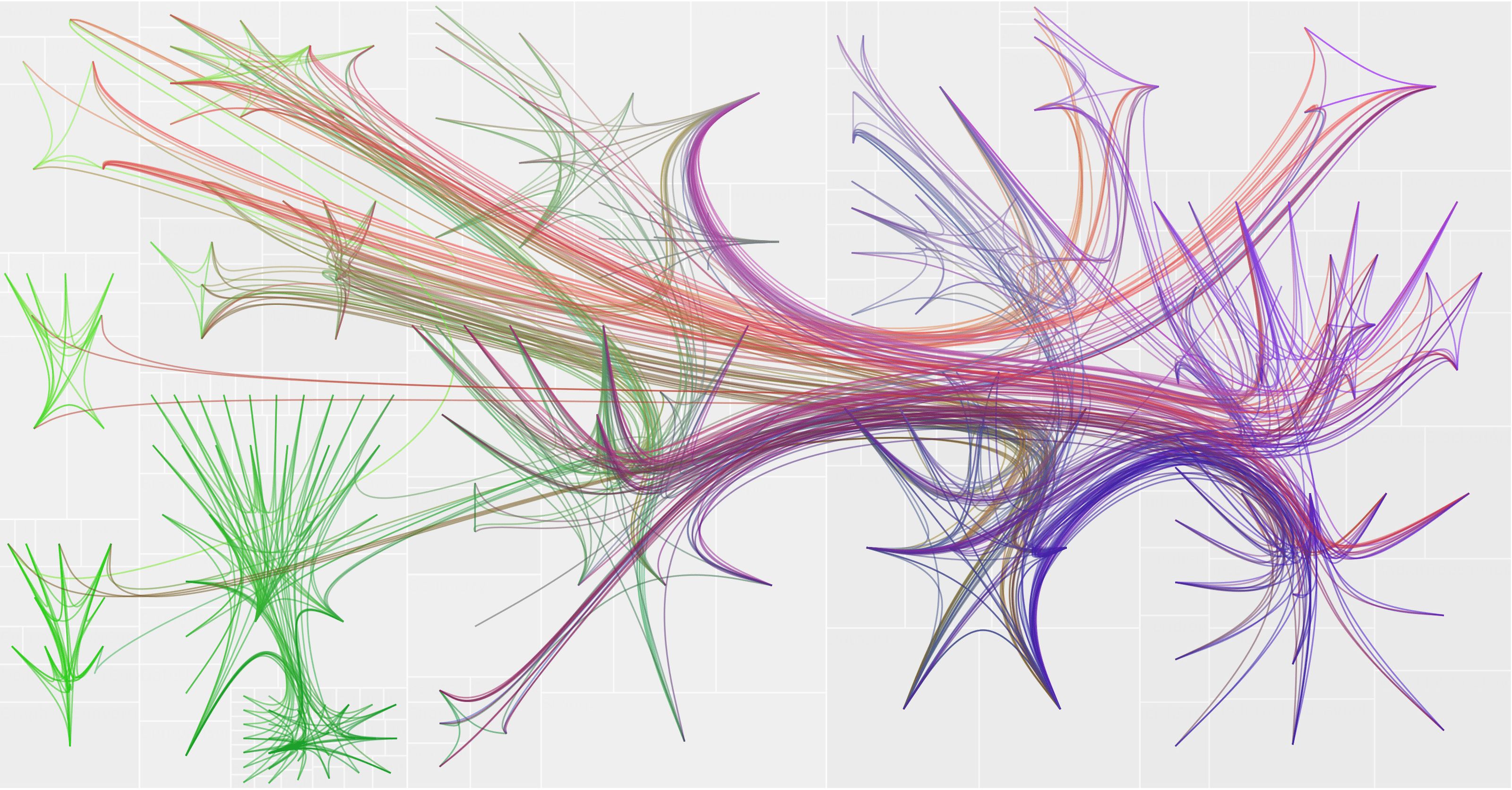}
\includegraphics[width=0.49\textwidth]{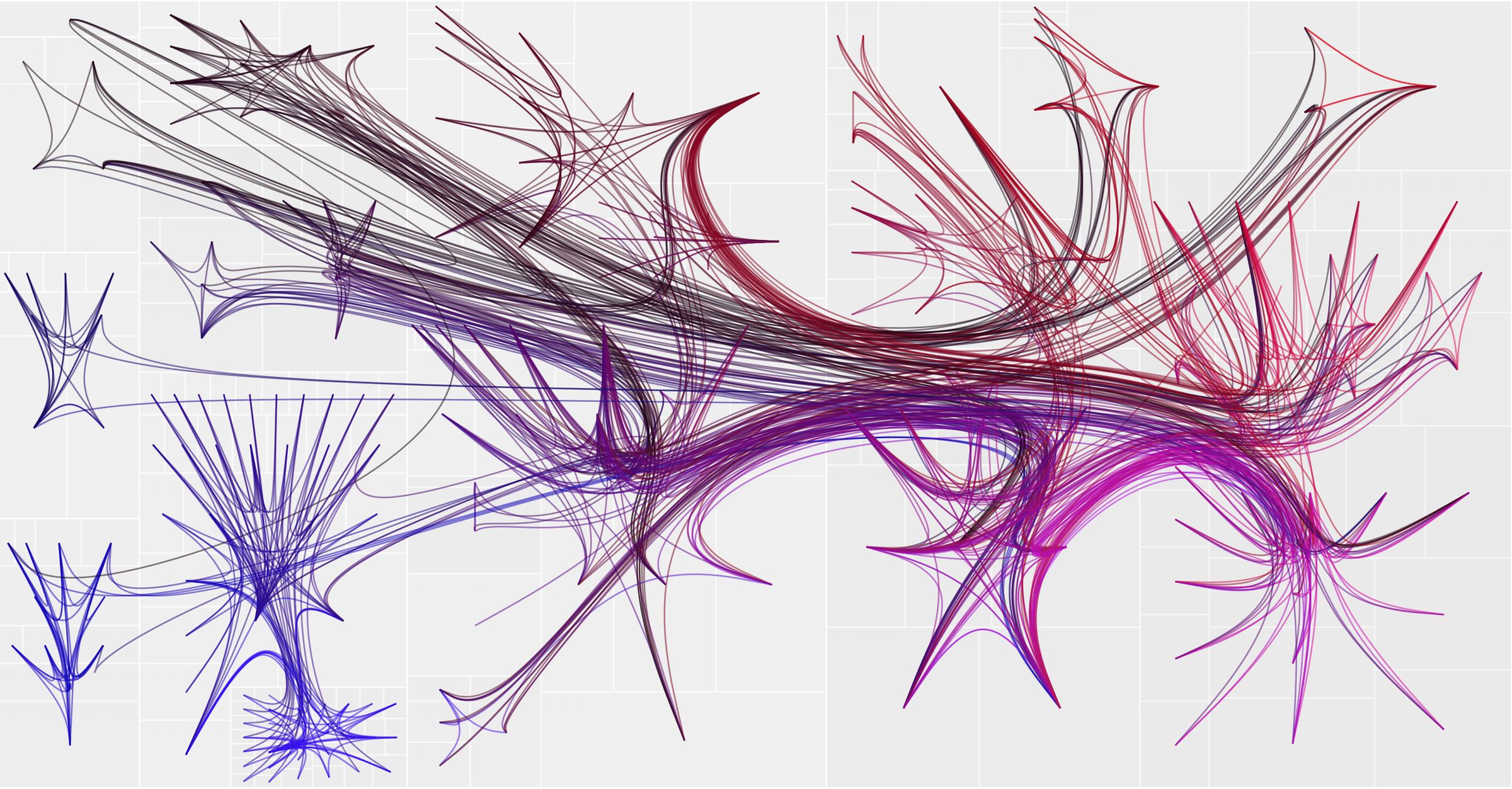}
\caption{Different coloring for ``tree map''}
\label{fig:treemap}
\end{subfigure}
\begin{subfigure}[t]{\textwidth}
\includegraphics[width=0.49\textwidth]{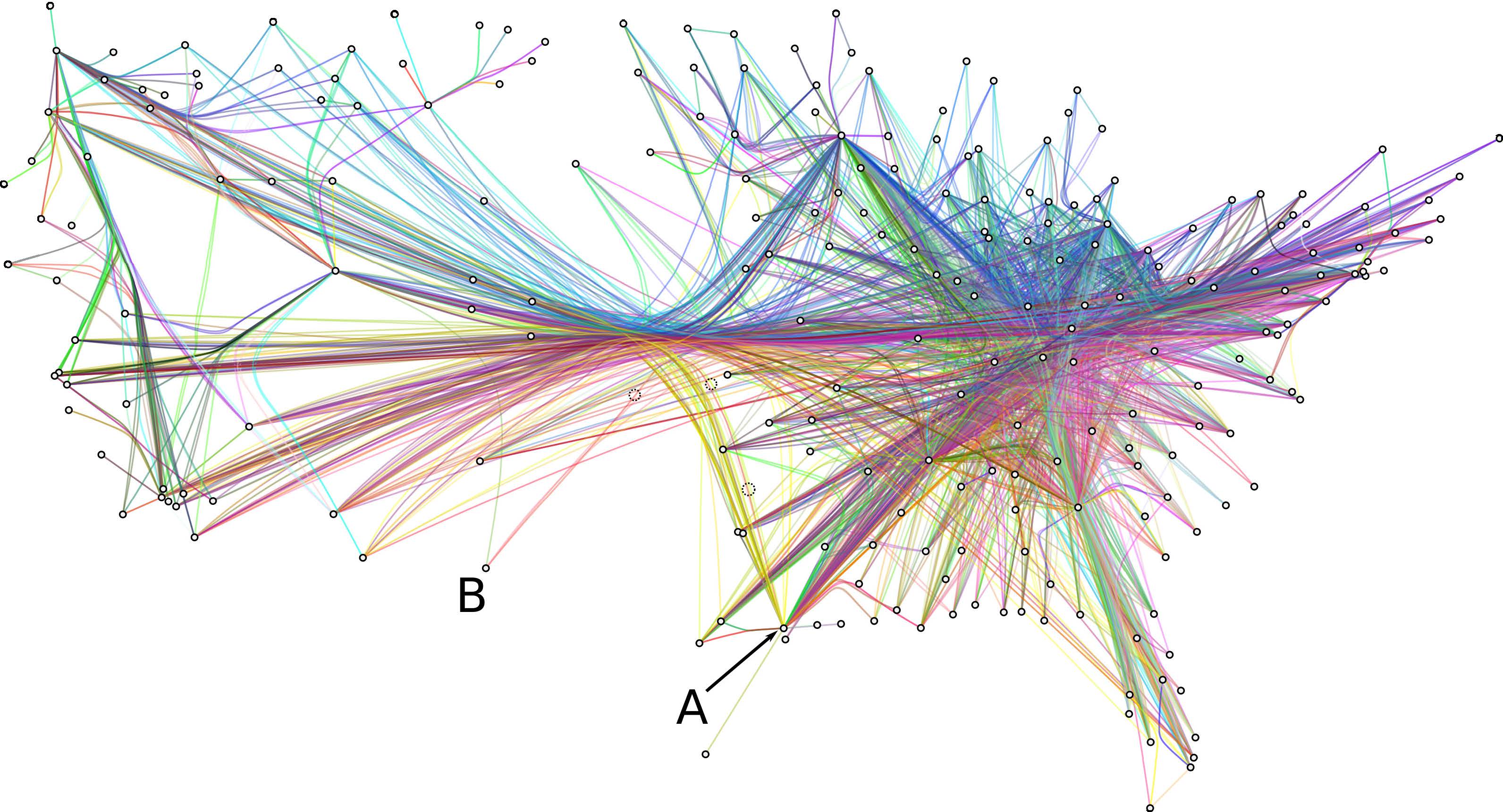}
\makebox[0.49\textwidth][c]{\includegraphics[width=0.25\textwidth]{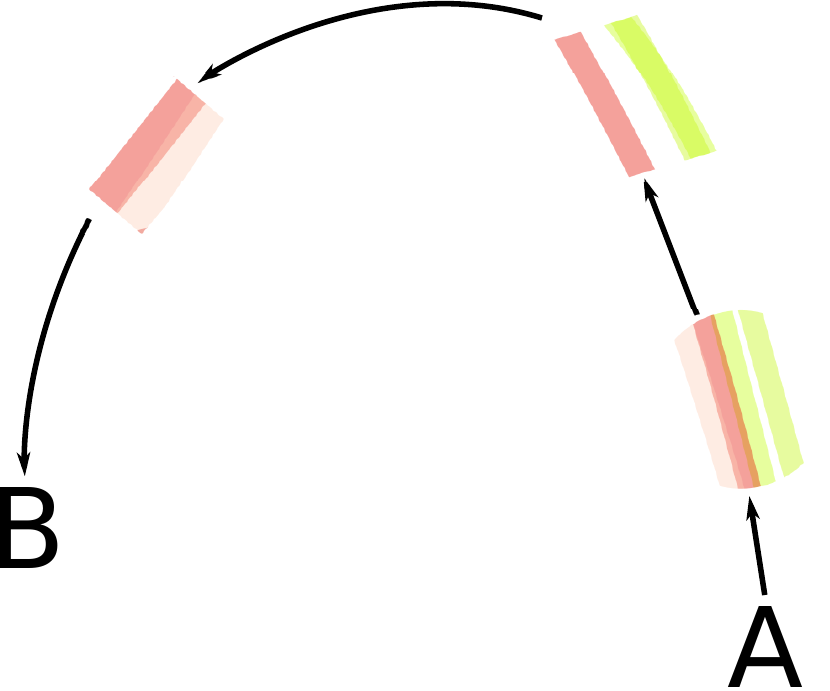}}
\includegraphics[width=0.49\textwidth]{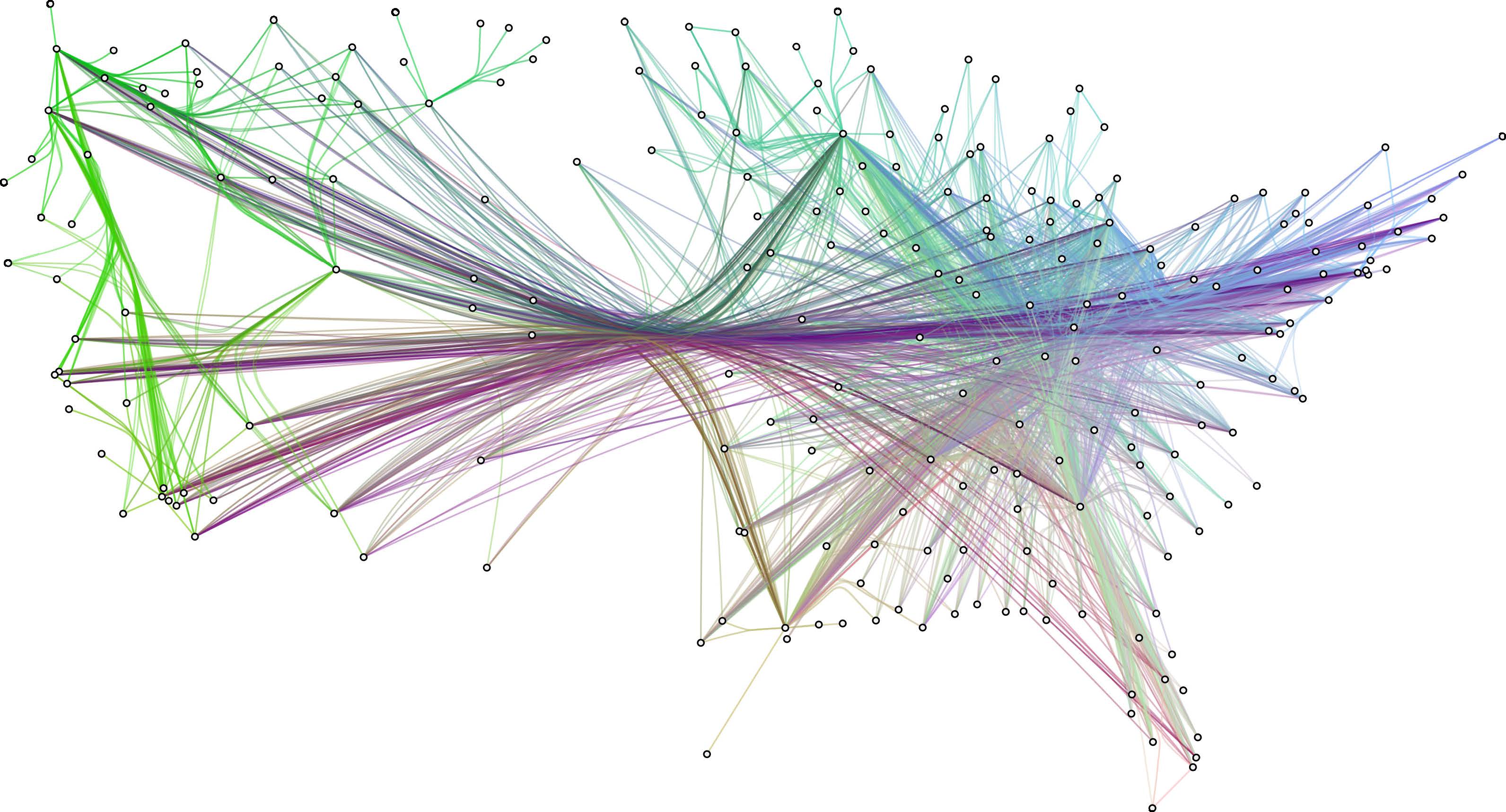}
\includegraphics[width=0.49\textwidth]{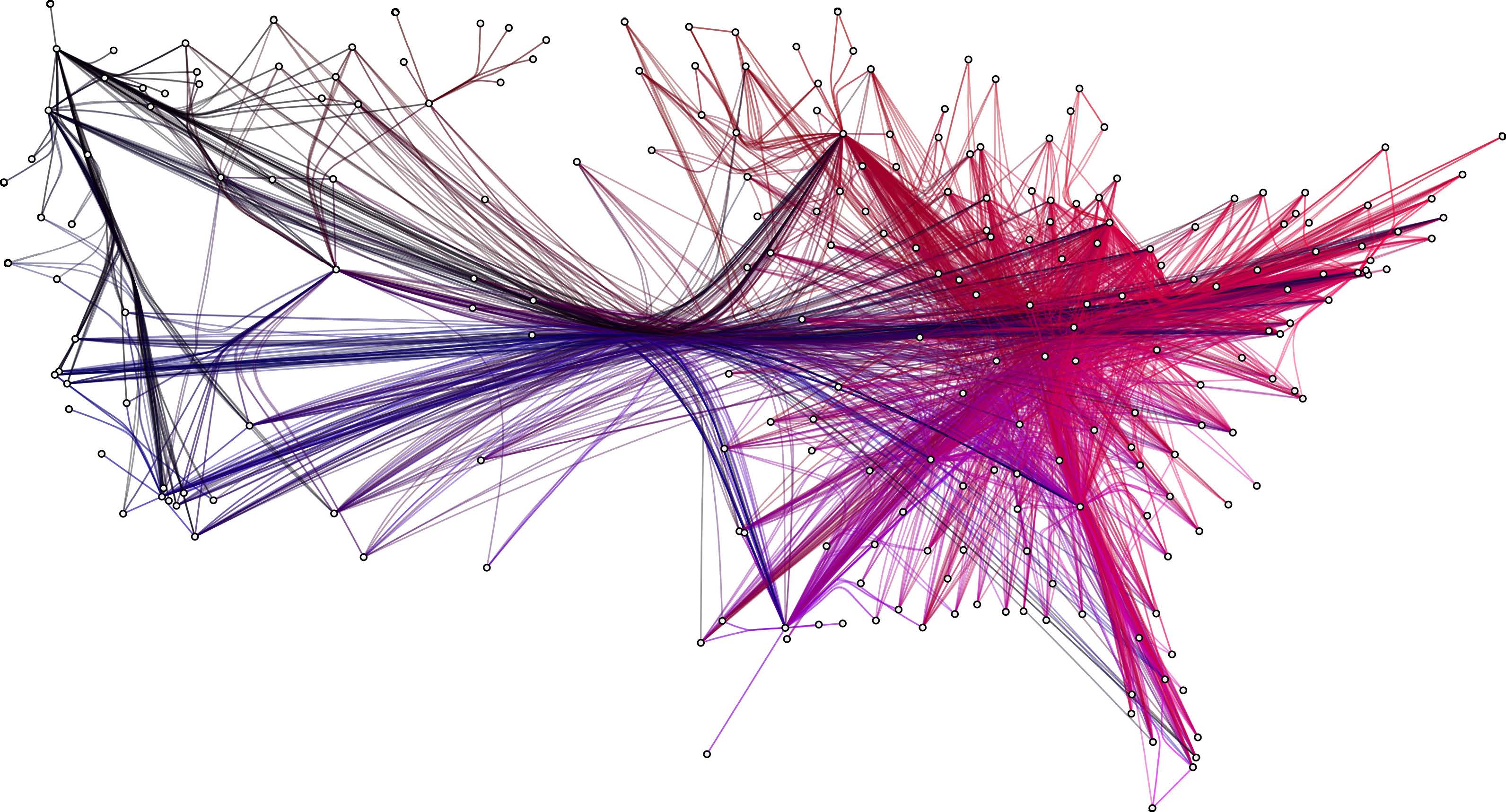}
\caption{Different coloring for ``airline''}
\label{fig:airline}
\end{subfigure}
\caption{Colorings for the graphs ``tree map'' and ``airline''. 
In both subfigures: \textbf{top left}: the colorings from Peacock with $\epsilon=0.001$. 
In Fig.~\ref{fig:treemap}, the local linear gradient is clearer at the the ends of the bundles. 
In Fig.~\ref{fig:airline}, the large bundle in the middle shows the local coloring, by separating the bundle into the upper blue dominating part, the middle red-ish part, and the lower lighter part.
\textbf{Top right}: examples of how the colorings enhance readability by investigating the parts within dashed line circles in both top-left subfigures.
In Fig.~\ref{fig:treemap}, the colors in bundle A help the user to recognize, for example, 
1) the blue-ish part in bundle A leads to the blue-ish part of the top-right ``claw'' or the right ``claw'';
2) the red-ish part in bundle A leads to the red-ish part of the ``claw'' at the right of bundle B, 
or the top right ``claw'';
3) the yellow-ish half that joins in the middle leads to bundle B or to the ``claw'' at the right of bundle B.
Fig.~\ref{fig:airline} shows how the coloring help a pink edge from A to B 
stand out against other edges in the same bundle. 
\textbf{Bottom left}: the coloring from Peacock with $\epsilon=1$. 
In Fig.~\ref{fig:treemap}, bundles are globally differentiated.
In Fig.~\ref{fig:airline}, for nodes of large degrees, 
the edges connecting to them have distinct colors for different directions.
\textbf{Bottom right}: the baseline coloring. 
In Fig.~\ref{fig:treemap}, the user may mis-recognize that there are edges from bundle A to bundle B.
In Fig.~\ref{fig:airline}, unlike the bottom-left subfigure, edges connecting to the same node tend to have similar colors.
}
\end{figure}

\begin{figure*}[!t]
\centering
\begin{tabular}{cc}
\includegraphics[trim={0 9cm 0 0},clip,width=0.49\textwidth]{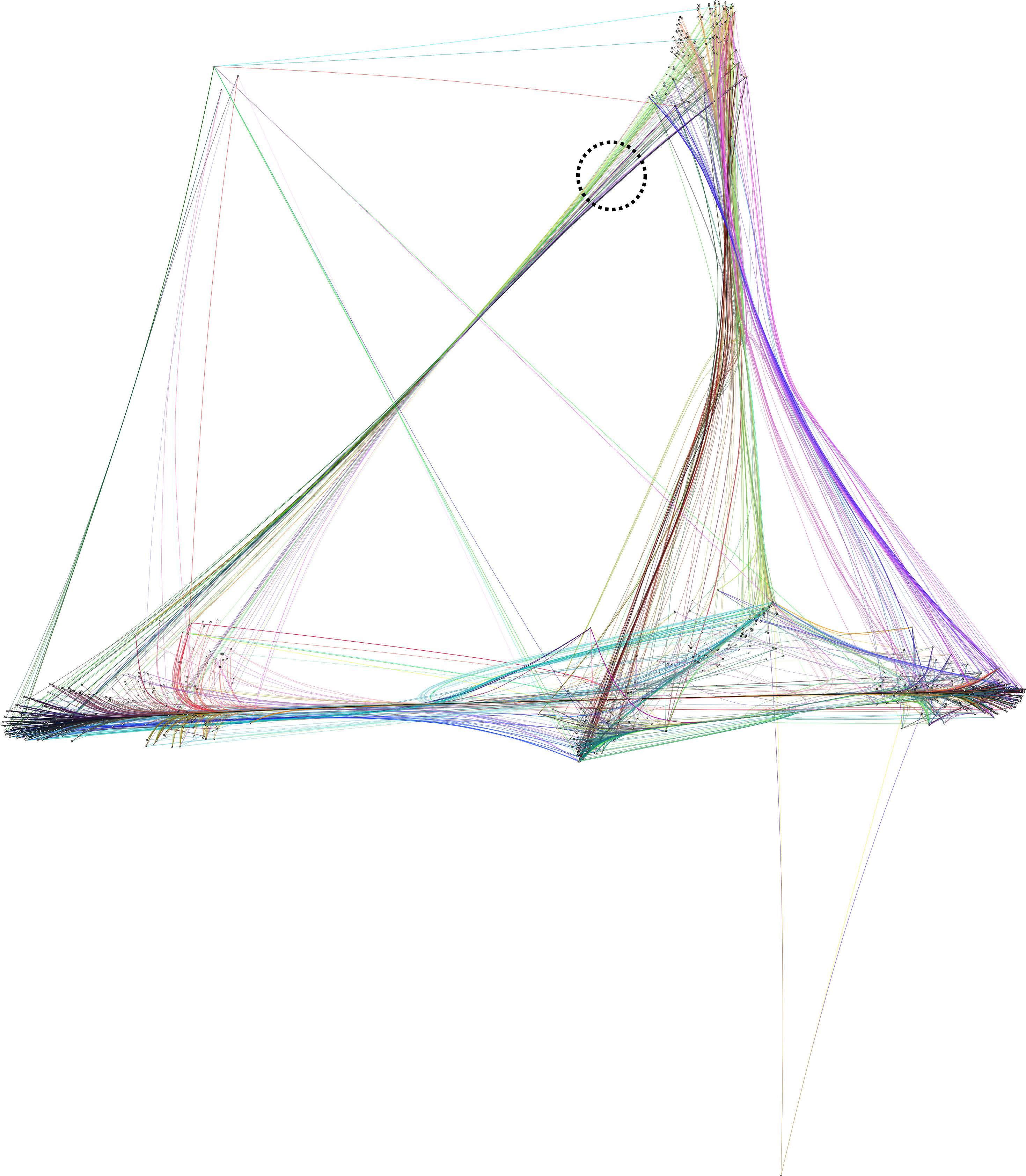}&
\includegraphics[width=0.3\textwidth]{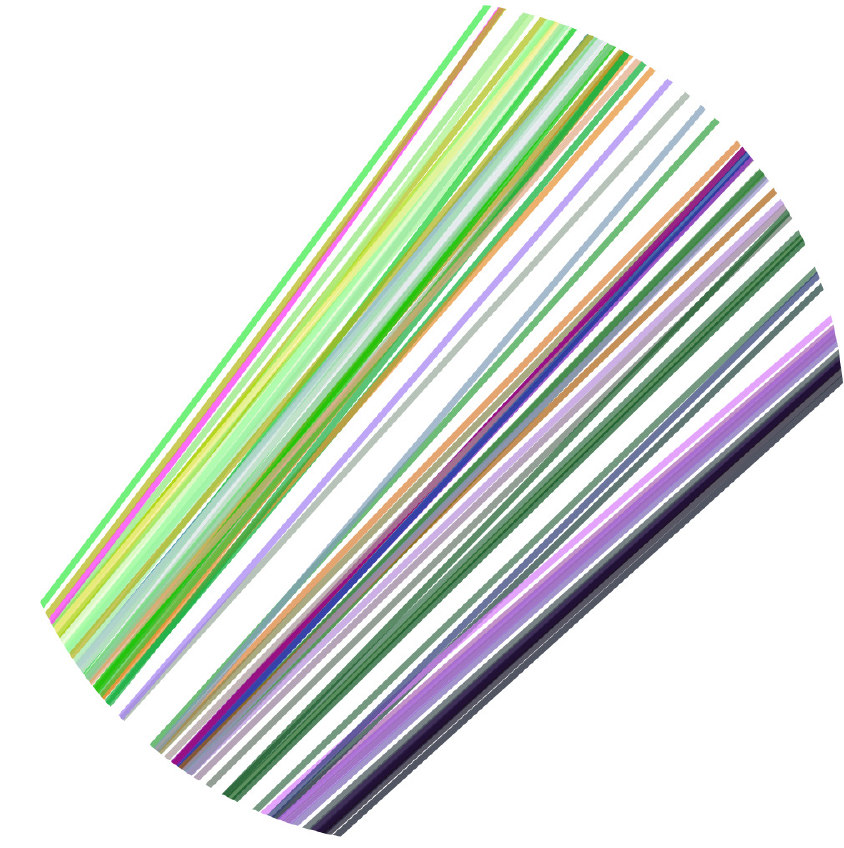}\\
\includegraphics[trim={0 9cm 0 0},clip,width=0.49\textwidth]{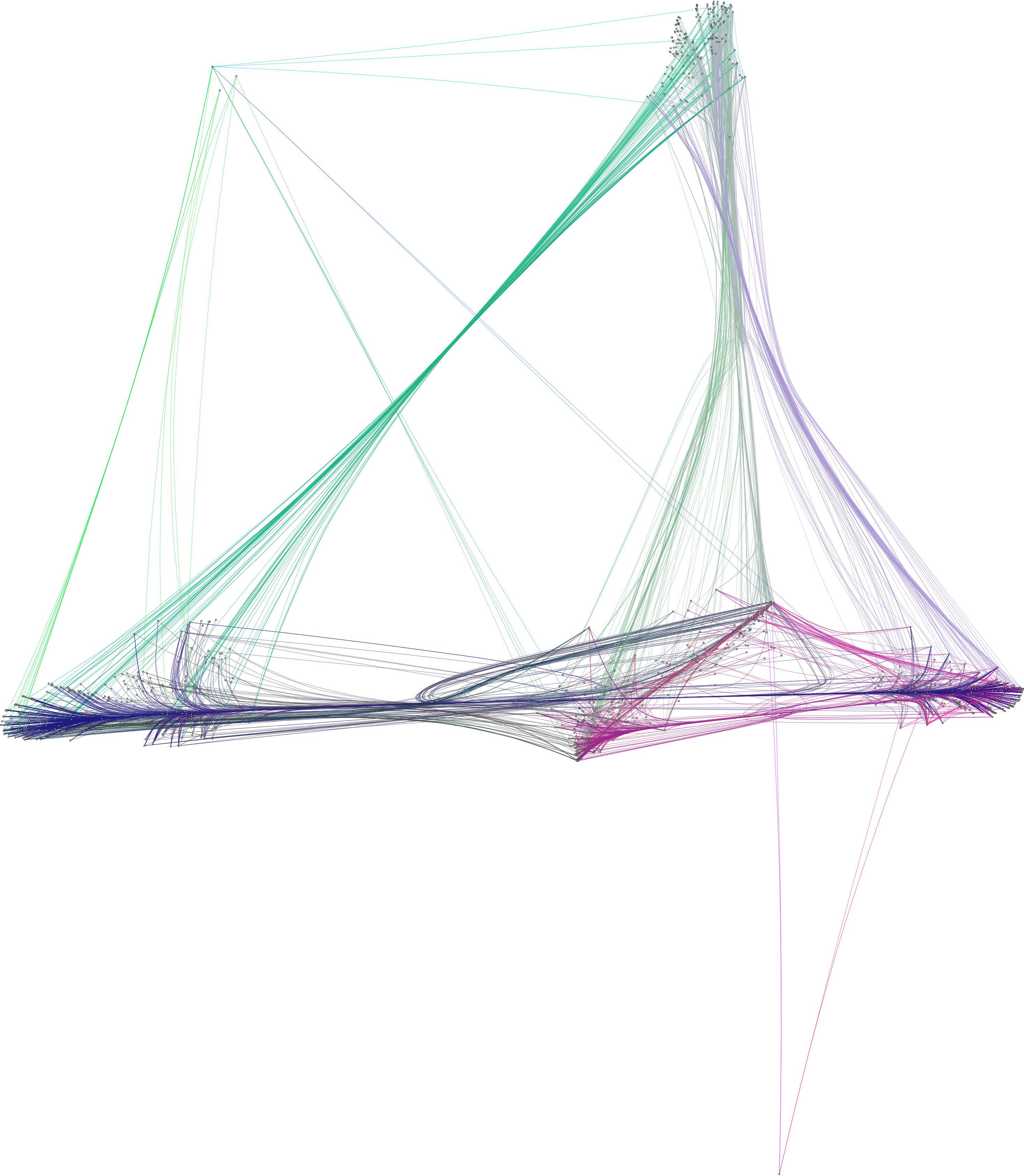}&
\includegraphics[trim={0 9cm 0 0},clip,width=0.49\textwidth]{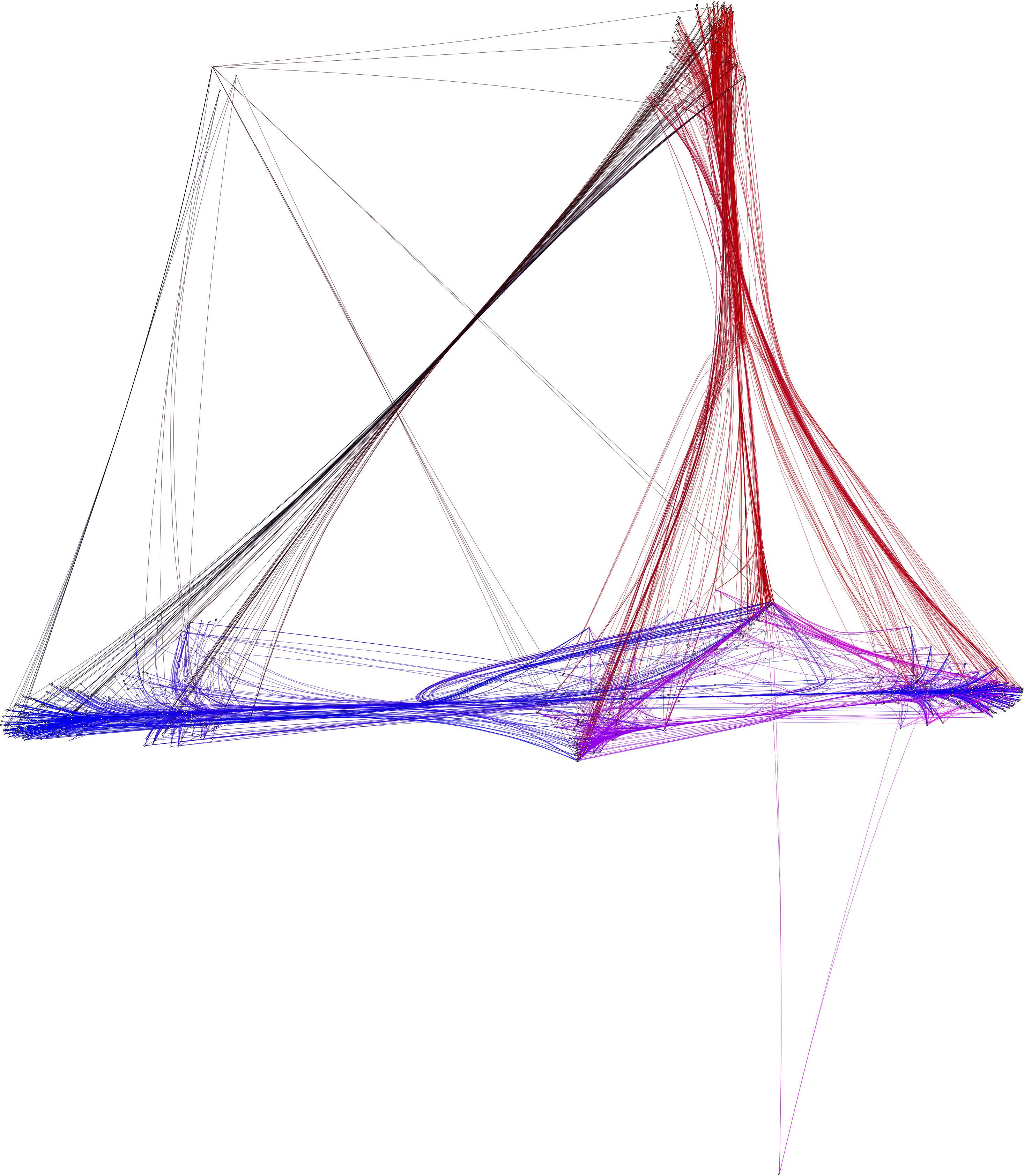}
\end{tabular}
\caption{Colorings for graph ``Jane Austen''. \textbf{Top left}: the coloring from Peacock with $\epsilon=0.001$. 
The ``crossing'' at the right half of the figure shows a clear differentiation: 
red edges go from upper-right to lower-left; 
green edges go from upper to lower; 
blue edges go from upper-left to lower-right. 
Without the local coloring, it is difficulty to tell 
whether the bundles or the edges are crossing or just tangental to one another.
\textbf{Top right}: another example from the zoomed-in version of the part 
within the dashed line circle in the top-left figure. 
Edge colors change from green to purple-ish. 
The colors help the user follow the edges after the heavily bundled part in the middle:
red edges mostly go leftwards, green edges are scattered, and blue edges mostly go rightwards.
\textbf{Bottom left}: the coloring from Peacock with $\epsilon=1$. 
Less locality but more globality. 
The earlier blue edges in the right ``crossing'' become purple, 
but still distinguishable from the other two bundles.
\textbf{Bottom right}: the baseline coloring, 
which loses the distinguishablity shown in the Peacock coloring.
}
\label{fig:jane-austen}
\end{figure*}

\begin{figure*}[!t]
\centering
\includegraphics[width=0.49\textwidth]{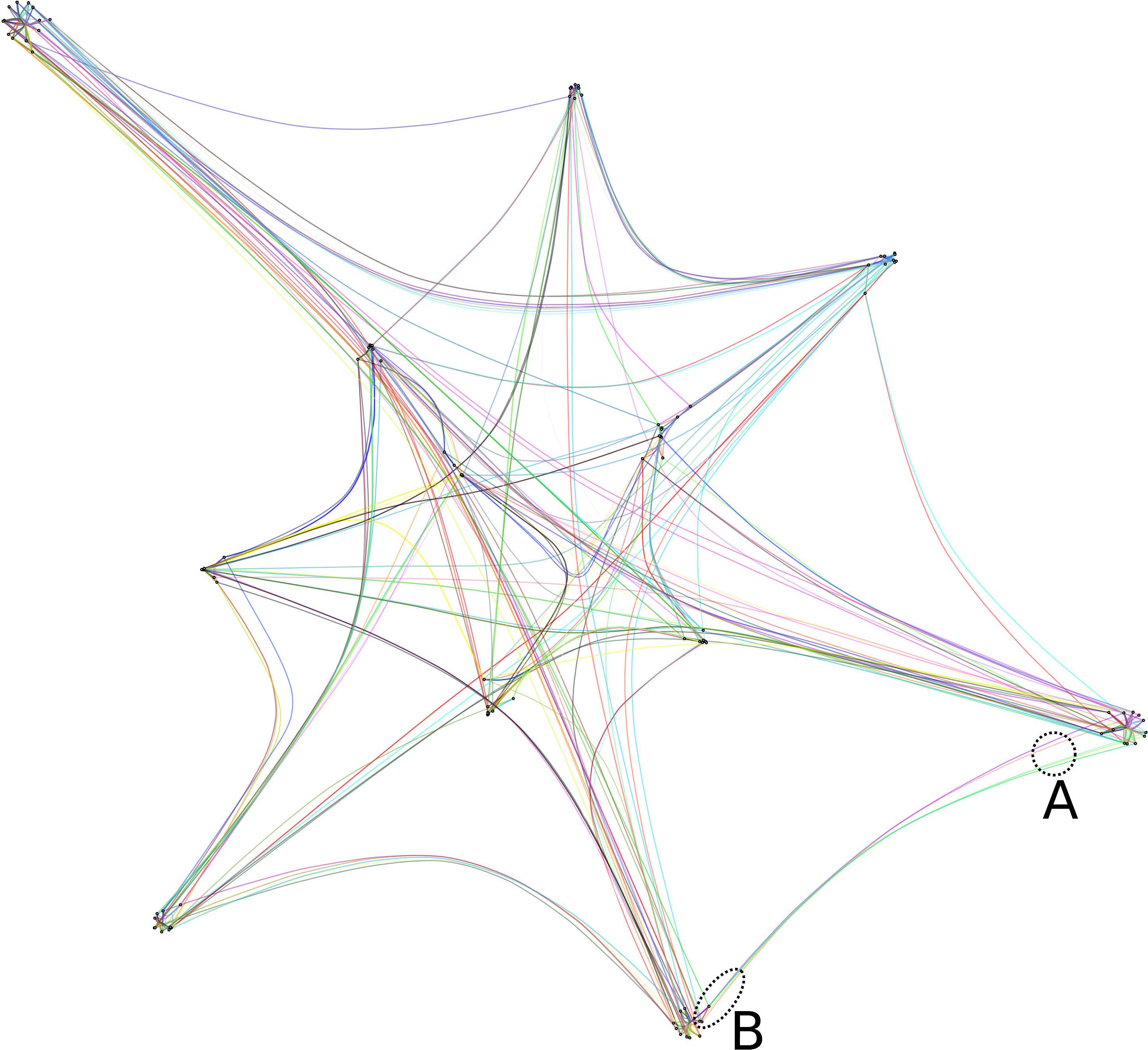}
\makebox[0.49\textwidth][c]{\includegraphics[width=0.35\textwidth]{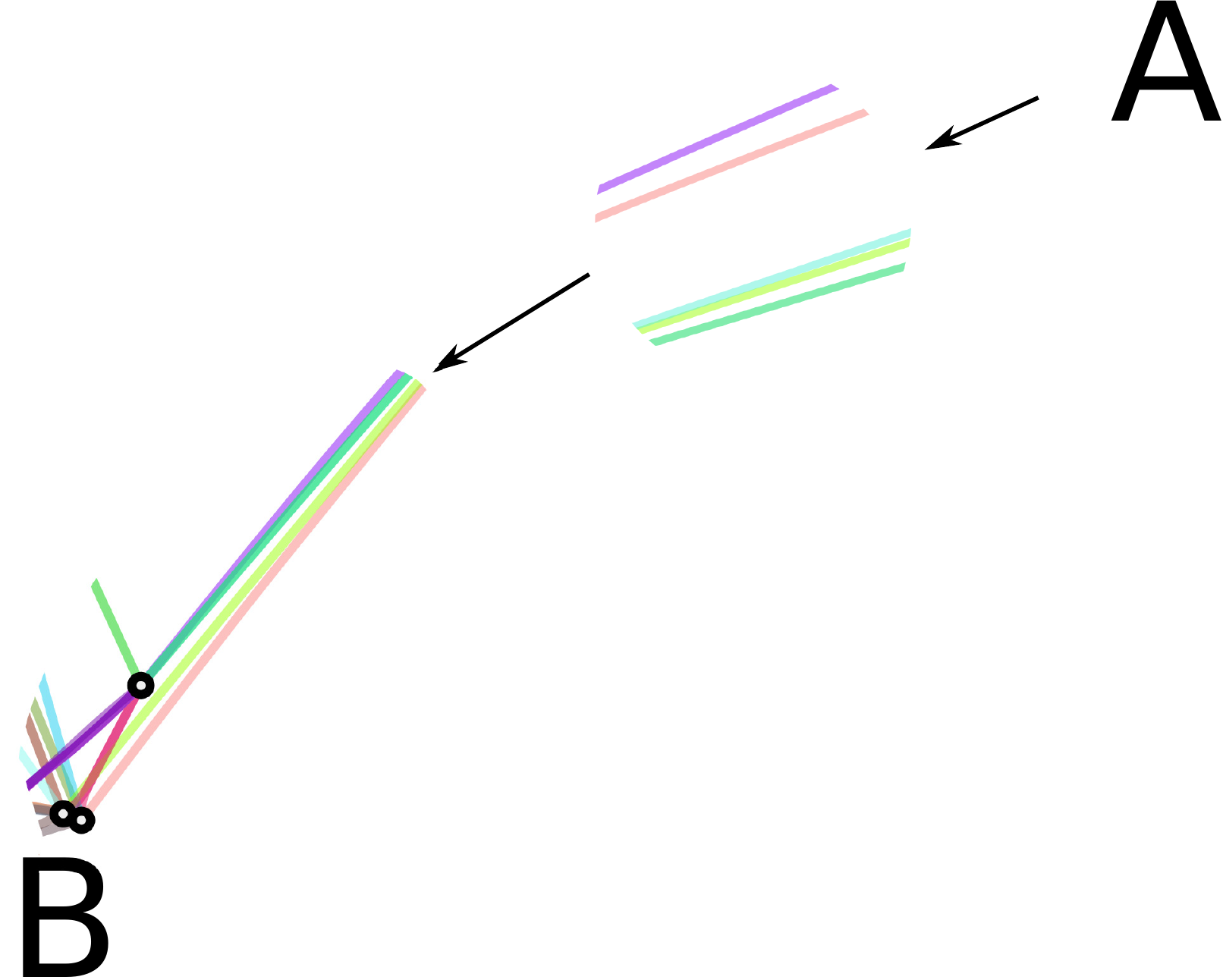}}
\includegraphics[width=0.49\textwidth]{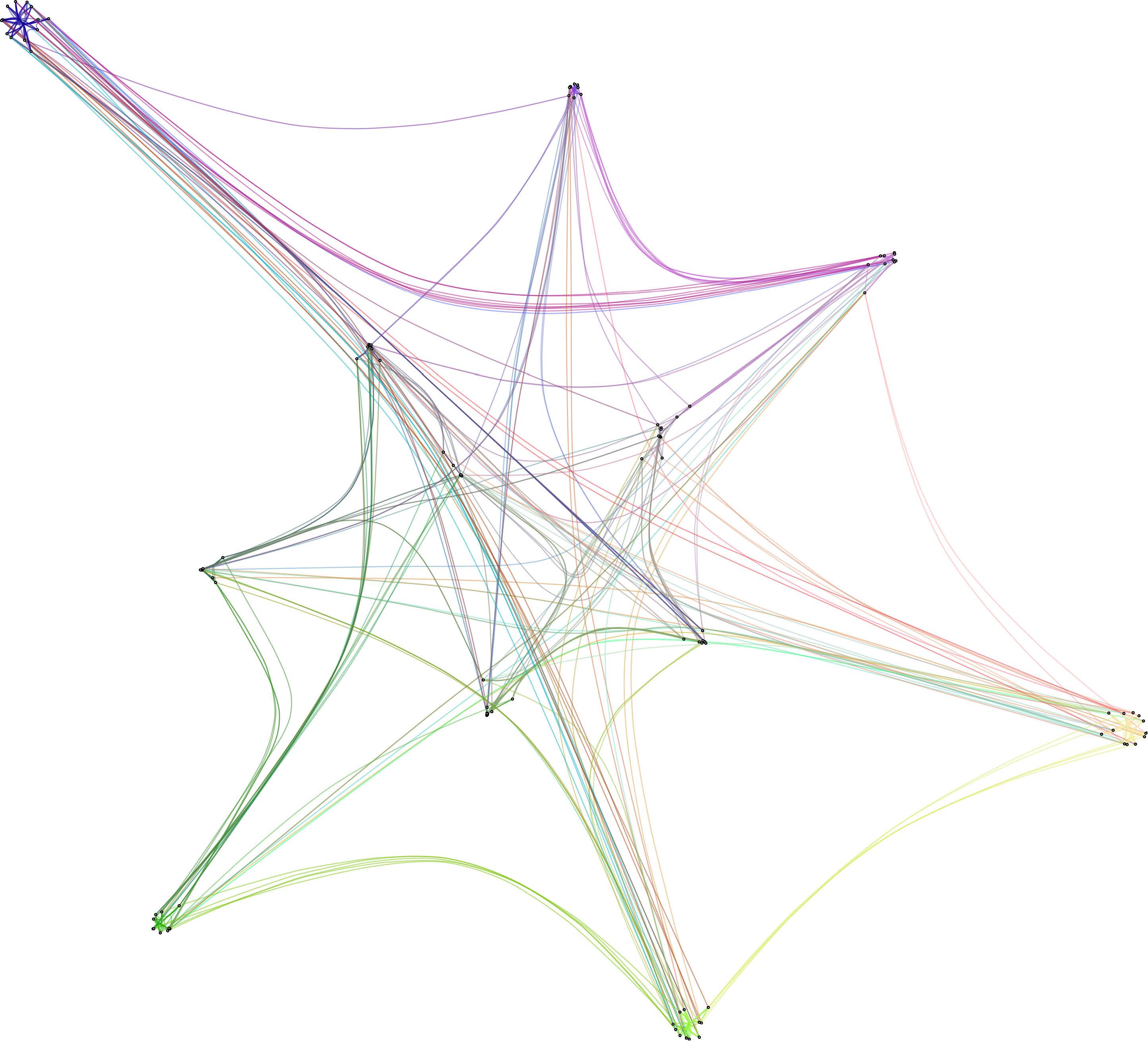}
\includegraphics[width=0.49\textwidth]{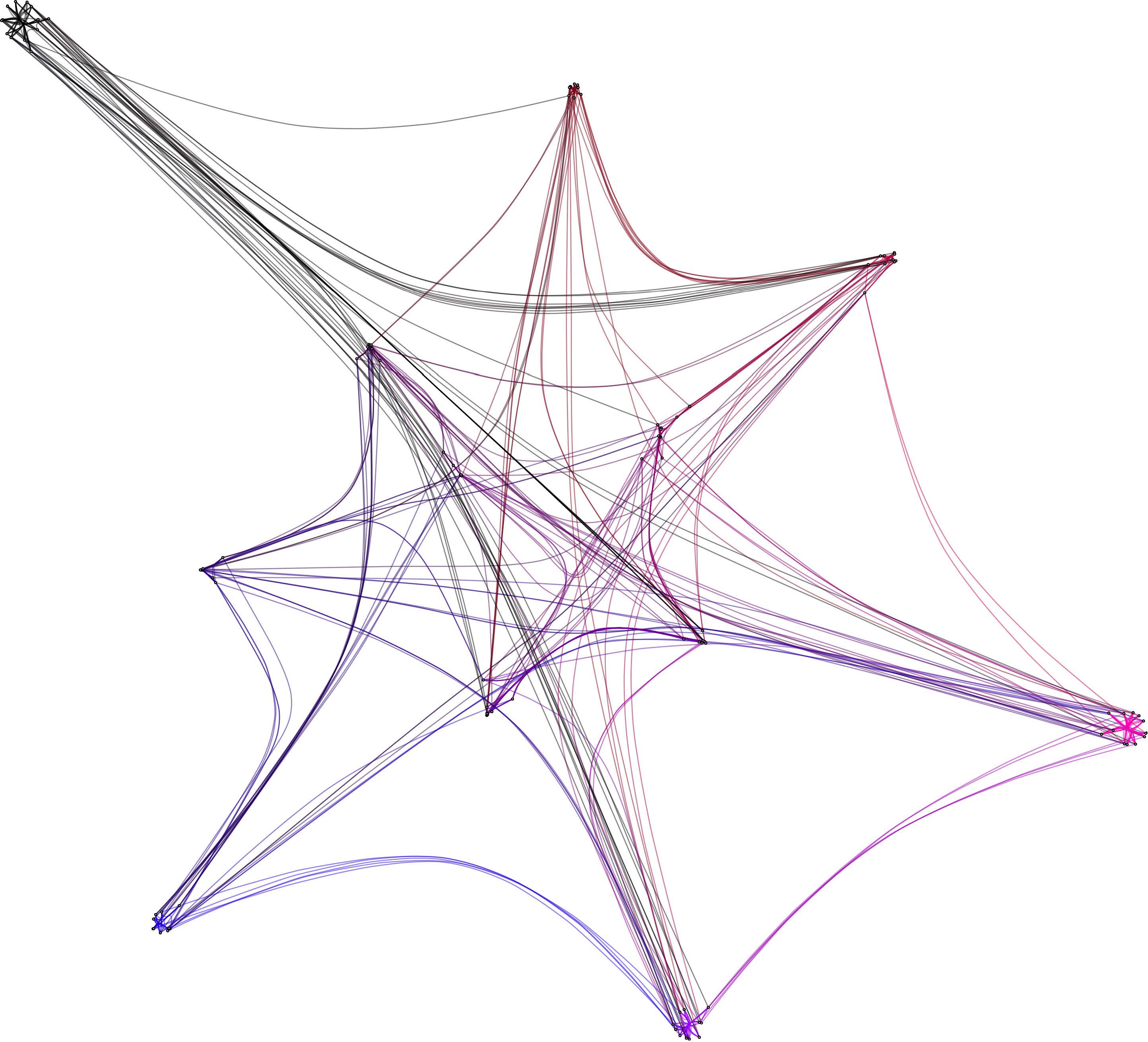}
\caption{Colorings for graph ``football''. 
textbf{Top left}: the coloring from Peacock with $\epsilon=0.001$. 
The uncertainty in this graph is mostly from the small clusters of nodes at the end of the edges.
\textbf{Top right}: an example showing how the coloring help distinguish 
heavily bundled edges from A to B.
The shown segments are the zoomed-in version of the parts within the 
dashed line circle or ellipse in the top-left figure.
We can see, for example, that the blue edge leads to the top node in B, 
while the yellow edge leads to leftmost node in B. 
This is also noticeable in the top-left figure, particularly for the blue edge.
However, it will be a difficult task with the baseline coloring 
since the the part between A and B is heavily bundled.
\textbf{Bottom left}: the coloring from Peacock with $\epsilon=1$. 
Colors of edges from the same cluster are differentiated (e.g., at the top-left cluster,
edge colors vary from red to blue).
\textbf{Bottom right}: the baseline coloring, which only reflects the locations of the bundles.
}
\label{fig:football}
\end{figure*}

We demonstrate the Peacock bundles method on five graphs (Figs. \ref{fig:circular}--\ref{fig:jane-austen}):
two graphs with hierarchical edge bundling \cite{holten06}, 
and three 
with force-directed edge bundling \cite{holten09}.
The two graphs with hierarchical edge bundling are created from the class hierarchy 
of the visualization toolkit Flare \cite{flare},
with the built-in radial layout (graph named ``radial''; Fig.~\ref{fig:circular}) and 
tree map layout (``tree map''; Fig.~\ref{fig:treemap}) in d3.js \cite{bostock11} respectively.
The four graphs with force-directed edge bundling are: 
a spatial graph of US flight connections (``airline''; Fig.~\ref{fig:airline}); 
a graph of consecutive word-to-word appearances in novels of Jane Austen (``Jane Austen''; Fig.~\ref{fig:jane-austen}); 
and
a graph of matches between US college football teams (``football''; Fig.~\ref{fig:football}).  
The last three graphs are laid out as an unconstrained 2D graph by a recent node-neighborhood preserving layout
method \cite{parkkinen10mlg}. For all graphs, edge bundles were created by a d3.js plugin 
implementing the algorithm \cite{d3-forcebundle} 
adapted to splines. 
All coloring are compared with a baseline coloring from end point positions.

\noindent\textbf{The baseline}. We compare our method with 
a baseline coloring that directly encodes end point positions into color channels. 
We choose channel red and blue for the encoding in the experiments.
Let $\mathbf{v}_i^1=(x_{i}^1, y_{i}^1)$ and $\mathbf{v}_i^2=(x_{i}^2, y_{i}^2)$ be the 
onscreen coordinates of edge $i$'s two end points as in \eqref{eq:dist-orig}. 
We first create a 3-dimensional vector $\widetilde{Col}^{\mathrm{baseline}}_i$ as the ``unnormalized'' color for edge $i$ as
\begin{equation}
\widetilde{Col}^{\mathrm{baseline}}_{i}=(\min(x_{i,1}, x_{i,2}), 0, \min(y_{i,1}, y_{i,2}))^\mathrm{T}
\end{equation}
then we affinely normalize the matrix $\widetilde{Col}^{\mathrm{baseline}}$ into $[0,1]$ 
to obtain the final baseline colors $Col^{\mathrm{baseline}}$.

\noindent\textbf{Choices of Peacock parameters}. The parameters $T$ and $K_{min}$ in 
\eqref{eq:thrsh-cp-int}
must
be chosen to determine $B_{ij}$. We set $T$ to  
$2\%\sim 4\%$ of $\max(\mbox{graph width}, \mbox{graph height})$, 
and fix $K_{min}$ as 0.4. 
Experiments show the choices give good results empirically.

Figures \ref{fig:circular} -- \ref{fig:jane-austen} show the results
from the proposed method and the baseline. The top-left subfigures are with the tradeoff parameter
set to prefer locality
in the coloring. The top-right subfigures provide zoomed-in views detailing
the local color variation (``peacock fans'') and demonstrating
how the coloring improves readability and helps follow edges.
The bottom-left figures are optimized to differentiate origins and destinations globally
(tradeoff parameter $\epsilon=1$), hence colors indicate overall trends of connections between
areas of the graph layout,
at the expense of less color variability within bundles.
The bottom-right figures are from the baseline, also aiming to show variability of endpoint positions 
the coloring but not optimized by machine learning; the simple baseline coloring leaves bundles 
and within-bundle variation less distinguishable.

\section{Conclusions}
We introduced ``peacock bundles'', a novel edge coloring algorithm for graphs with edge bundling.
Colors are optimized both to preserve differences between bundle locations, 
and differentiate edges within bundles.
The algorithm is based on dimensionality reduction
without need to explicitly define bundles.
Experiments show the method
outperforms the baseline
coloring with several graphs and bundling algorithms,
greatly improving the comprehensibility of graphs with edge bundling.
Potential future work includes incorporating color perception models
\cite{jeffrey2012colornaming}, and more nuanced weighting schemes for global-local tradeoffs.

We acknowledge computational resources from the Aalto Science-IT project.
Authors belong to the COIN centre of excellence.
The work was supported by Academy of Finland grants 252845 and 256233.

\clearpage
\bibliographystyle{splncs03}
\bibliography{bundlecoloring}

\end{document}